\documentclass[12pt]{article}

\usepackage{latexsym}
\usepackage{amssymb,amsfonts,amsmath}
\usepackage{graphicx} 
\usepackage{indentfirst}
\usepackage{bbm}
\usepackage{amssymb}
\usepackage{verbatim}
\usepackage{amsmath, amsthm,amssymb}
\usepackage{mathrsfs}
\usepackage{hyperref}
\usepackage{amsfonts}
\usepackage{dsfont}
\usepackage{cite}
\usepackage{xcolor}
\usepackage{slashed}
\PassOptionsToPackage{linecolor=blue,backgroundcolor=blue!25,bordercolor=blue,textsize=scriptsize}{todonotes}
\usepackage{todonotes}

\parskip=\medskipamount

\usepackage[mathscr]{euscript}

\newcommand{\de}{{\nabla}}

\numberwithin{equation}{section}

\newcommand {\cD}{{\cal D}}

\newcommand {\cG}{{\cal G}}
\newcommand {\cH}{{\cal H}}

\newcommand {\cJ}{{\cal J}}
\newcommand {\cK}{{\cal K}}
\newcommand {\cL}{{\cal L}}

\newcommand {\cN}{{\cal N}}
\newcommand {\cO}{{\cal O}}

\newcommand {\cR}{{\cal R}}

\newcommand{\bbD}{\mathbb D}

\def\a{\alpha}
\def\b{\beta}

\def\d{\delta}
\def\e{\epsilon}

\def\g{\gamma}
\def\G{\Gamma}

\def\k{\kappa}
\def\l{\lambda}

\def\o{\omega}

\def\q{\theta}

\def\s{\sigma}

\def\D{\Delta}
\def\F{\Phi}

\def\L{\Lambda}
\def\O{\Omega}

\def\S{\Sigma}

\def\ri{{\rm i}}

\newcommand{\rd}{\mathrm d}
\newcommand{\hm}{{{m}}}
\newcommand{\hn}{{{n}}}
\newcommand{\hp}{{{p}}}
\newcommand{\hq}{{{q}}}
\newcommand{\ha}{{{a}}}
\newcommand{\hb}{{{b}}}
\newcommand{\hc}{{{c}}}
\newcommand{\hd}{{{d}}}
\newcommand{\he}{{{e}}}
\newcommand{\hM}{{{M}}}
\newcommand{\hN}{{{N}}}
\newcommand{\hA}{{{A}}}
\newcommand{\hB}{{{B}}}
\newcommand{\hC}{{{C}}}
\newcommand{\hal}{{{\a}}}
\newcommand{\hbe}{{{\b}}}
\newcommand{\hga}{{{\g}}}
\newcommand{\hde}{{{\d}}}
\newcommand{\hrh}{{{\rho}}}

\newcommand{\ve}{\varepsilon}

\newcommand{\pa}{\partial}
\newcommand{\hf}{\frac12}

\newcommand{\vf}{\varphi}

\newcommand{\be}{\begin{equation}}
\newcommand{\ee}{\end{equation}}
\newcommand{\bea}{\begin{eqnarray}}
\newcommand{\eea}{\end{eqnarray}}
\newcommand{\non}{\nonumber}
\newcommand{\ba}{\begin{array}}
\newcommand{\ea}{\end{array}}

\newcommand{\bsubeq}{\begin{subequations}}
\newcommand{\esubeq}{\end{subequations}}

\newcommand{\bm}[1]{\mbox{\boldmath$#1$}}

\def\double #1{#1{\hbox{\kern-2pt $#1$}}}

\newcommand{\eps}{\varepsilon}

\newcommand{\eol}{\notag \\}

\newcommand{\loco}{\vert}

\expandafter\ifx\csname url\endcsname\relax
  \def\url#1{\texttt{#1}}\fi
\expandafter\ifx\csname urlprefix\endcsname\relax\fi
\providecommand{\eprint}[2][]{\url{#2}}
\begin{document}
%%%%%%%%%%%%%%%%
%%%%%%%%%%%%%%%%

\begin{titlepage}
\begin{flushright}
March, 2023
\end{flushright}
\vspace{5mm}

\begin{center}
{\Large \bf 
On curvature-squared invariants of minimal five-dimensional supergravity from superspace
}
\end{center}

\begin{center}

{\bf
Gregory Gold,
Jessica Hutomo,
Saurish Khandelwal,\\
and Gabriele Tartaglino-Mazzucchelli
} \\
\vspace{5mm}

\footnotesize{
{\it 
School of Mathematics and Physics, University of Queensland,
\\
 St Lucia, Brisbane, Queensland 4072, Australia}
}
\vspace{2mm}
~\\
\texttt{g.gold@uq.edu.au; 
j.hutomo@uq.edu.au; 
s.khandelwal@uq.edu.au;
\\ g.tartaglino-mazzucchelli@uq.edu.au}\\
\vspace{2mm}

\end{center}

\begin{abstract}
\baselineskip=14pt

We elaborate on the off-shell superspace construction of curvature-squared invariants in minimal five-dimensional supergravity. This is described by the standard Weyl multiplet of conformal supergravity coupled to two compensators being a vector multiplet and a linear multiplet. In this set-up, we review the definition of the off-shell two-derivative gauged supergravity together with the three independent four-derivative superspace invariants defined in  arXiv:1410.8682. We provide the explicit expression for the linear multiplet based on a prepotential given by the logarithm of the vector multiplet primary superfield. We then present for the first time the primary equations of motion for minimal gauged off-shell supergravity deformed by an arbitrary combination of these three four-derivative locally superconformal invariants. We also identify a four-derivative invariant based on the linear multiplet compensator and the kinetic superfield of a vector multiplet which can be used to engineer an alternative supersymmetric completion of the scalar curvature squared.

\end{abstract}
\vspace{5mm}

\vfill
\end{titlepage}

%%%%%%%%%%%%

\newpage
\renewcommand{\thefootnote}{\arabic{footnote}}
\setcounter{footnote}{0}

\tableofcontents{}
\vspace{1cm}
\bigskip\hrule

\allowdisplaybreaks
%%%%%%%%%%%%%%%%%%%%%%%%%%%%%%%%%
\section{Introduction}

Almost  five decades  after the first (two-derivative) supergravity was constructed (for $\cN=1$ supersymmetry in four dimensions), higher-order locally supersymmetric invariants are still largely unknown. Higher-order curvature terms play, however, a significant role  in string theory,  where quantum corrections take the form of an infinite series potentially constrained by supersymmetry order by order in the string tension $\alpha'$ and the string coupling $g_s$. Many open problems in string theory, for example its vacua structure, are unresolved due to the lack of information about the full quantum corrected supergravity effective action. The complexity of such an effective theory is even made worse by the fact that the purely gravitational higher-curvature terms are related by supersymmetry to contributions depending on $p$-forms, which describe part of the string spectrum. These terms, which have not yet been systematically understood, play an important role in studying, for example, the moduli in compactified string theory and the low-energy description of string dualities; see, e.\,g., \cite{Antoniadis:1997eg,Antoniadis:2003sw,Liu:2013dna}. In the context of string-inspired holographic dualities, such as the AdS/CFT, higher-order $1/N$ corrections in quantum field theories translate into higher curvature terms on the gravity side making these contributions fundamental for precision tests in AdS/CFT. New interesting analyses in this topic have been performed in the last few years -- see for example \cite{Baggio:2014hua,Bobev:2020zov,Bobev:2020egg,Bobev:2021oku,Bobev:2021qxx,Liu:2022sew,Hristov:2022lcw,Bobev:2022bjm,Cassani:2022lrk} 
and references therein.

One obstacle to constructing locally supersymmetric higher-order invariants is that often supersymmetry is only realised {\emph{on-shell}}, meaning {\emph{the symmetry algebra closes by using equations of motion}}. In on-shell approaches -- which are, e.g., typically used in 10- and 11-dimensional theories -- one needs to intertwine the construction of higher-order invariant terms in the Lagrangian of interest with a systematic and consistent deformation of the supersymmetry transformations making the problem remarkably involved. 
This obstacle is simplified by using {\emph{off-shell supersymmetry}} where one introduces extra (auxiliary) degrees of freedom to obtain supersymmetric multiplets possessing {\emph{model-independent transformation rules}}.  In a low number of space-time dimensions (D), in particular ${\rm D}\leq 6$, off-shell techniques are by now well developed and understood for up to eight real supercharges -- see \cite{FVP,Lauria:2020rhc,SUPERSPACE,WB,BK, Kuzenko:2022skv,Kuzenko:2022ajd} for reviews on off-shell approaches to supersymmetry and supergravity. In these cases, the construction of supergravity higher-derivative invariants can in principle be systematically approached. A restricted list of references using off-shell approaches to construct locally supersymmetric higher-derivative invariants is:
\cite{HOT,BSS1,LopesCardoso:1998tkj,Mohaupt:2000mj,BRS,CVanP,Bergshoeff:2012ax,Butter:2013rba,Butter:2013lta,Kuzenko:2013vha,OP1,OP2,OzkanThesis,BKNT-M14,Kuzenko:2015jxa,BKNT16,Butter:2016mtk,BNT-M17,NOPT-M17,Butter:2018wss,Butter:2019edc,Hegde:2019ioy,Mishra:2020jlc,deWS,Ozkan:2016csy,deWit:2010za,Butter:2014iwa}.

The scope of our paper is to enhance the classification of off-shell curvature-squared invariants of minimal five-dimensional (5D) supergravity. 
Minimal on-shell 5D supergravity was introduced four decades ago in \cite{Cremmer,CN}, and the first off-shell description was given in \cite{Howe5Dsugra} by the use of superspace techniques. Since then, 5D minimal supergravity  and its matter couplings have been extensively studied at the component level, both  in on-shell \cite{GST1,GST2,GZ,CD} and off-shell \cite{Zucker1,Zucker2,Zucker3,Ohashi1,Ohashi2,Ohashi3,Ohashi4,Bergshoeff1,Bergshoeff2,Bergshoeff3} settings. 
The superspace approach to general off-shell 5D $\cN=1$ supergravity-matter systems has then been developed in \cite{KT-M_5D2,KT-M_5D3,KT-M08,BKNT-M14}; see also \cite{Howe:2020hxi} for a recent local supertwistor description of 5D conformal supergravity.

In our paper we will specifically use the 5D $\cN=1$ conformal superspace approach of \cite{BKNT-M14}.\footnote{Conformal superspace was originally introduced by D.\,Butter for 4D $\cN=1$ supergravity in \cite{Butter4DN=1} and then extended to other space-time dimensions $2\leq {\rm D} \leq 6$ for various amount of supersymmetry in \cite{Butter4DN=2,BKNT-M1,BKNT-M14,BKNT16,BNT-M17,Kuzenko:2022qnb} -- see also \cite{Kuzenko:2022skv,Kuzenko:2022ajd} for recent reviews.} This approach merges advantages of the 5D superconformal tensor calculus of \cite{Ohashi1,Ohashi2,Ohashi3,Ohashi4,Bergshoeff1,Bergshoeff2,Bergshoeff3} with the superspace approaches of \cite{Howe5Dsugra,KT-M_5D2,KT-M_5D3,KT-M08}. 
 In the superconformal  setup (both in components and superspace), one enlarges the supergravity gauge group to be described by local superconformal transformations, plus potentially  internal symmetries. Local Poincar\'e supersymmetry is then recovered by using an appropriate choice of compensating multiplets that are used to gauge fix extra non-physical symmetries within the conformal algebra. For instance, in this setup, the off-shell formulation of minimal 5D supergravity is achieved by coupling the standard Weyl multiplet of 5D conformal supergravity to two off-shell conformal compensators: a vector multiplet and a hypermultiplet, the latter conveniently described by a linear multiplet. These will be the off-shell multiplets used in our paper. Within this setup, locally supersymmetric completions of the Weyl tensor squared and the scalar curvature squared were constructed for the first time, respectively, in \cite{HOT} and \cite{OP2} by using component fields techniques.
 Up to total derivatives, a generic combination of curvature-squared terms in 5D should include also a Ricci tensor squared invariant. A third independent locally superconformal invariant which includes Ricci squared was indeed constructed in superspace in \cite{BKNT-M14} by using a 5D analog of the ``log multiplet'' construction in 4D $\cN=2$ supergravity of \cite{Butter:2013lta}. However, due to the computational complexity of the log multiplet  in 5D, the component analysis of this invariant has not appeared so far -- in a follow-up paper, we will report on the component structure of this invariant which has been computed by making use of the computer algebra program {\it Cadabra} \cite{Cadabra-1,Cadabra-2}. 

Note that the conformal approach described above is not unique. In five dimensions it is known that an efficient setup to describe general supergravity-matter couplings make use of a vector-dilaton Weyl multiplet as a multiplet of conformal supergravity in place of the standard Weyl one \cite{Ohashi3, Bergshoeff1}.\footnote{The vector-dilaton Weyl multiplet terminology is used here to stress that the variant multiplet of conformal supergravity in \cite{Ohashi3, Bergshoeff1} is defined as an on-shell vector multiplet coupled to the standard Weyl multiplet. It was recently shown in \cite{Gold:2022bdk,Hutomo:2022hdi} that an on-shell hypermultiplet in a standard Weyl multiplet background can be reinterpreted as yet another new variant Weyl multiplet of off-shell conformal supergravity which was referred to as hyper-dilaton Weyl.}
A remarkable property of systems based on the use of a 5D vector-dilaton Weyl multiplet, that are related to the Poincar\'e supergravity first introduced in \cite{NR}, is the simplicity to define a third locally supersymmetric curvature-squared invariant. In fact, by employing a map between fields of the vector-dilaton Weyl multiplet and an off-shell vector multiplet, in \cite{BRS} a locally supersymmetric extension of the Riemann tensor squared was constructed (a construction that however is not applicable for a standard Weyl multiplet). This, together with the Weyl-squared invariant of \cite{HOT}, was sufficient for Ozkan \& Pang to construct in  \cite{OP1,OP2} a locally supersymmetric extension of the Gauss-Bonnet combination which is expected to play a key role in the description of the first $\alpha^\prime$ corrections to compactified string theory \cite{Zwiebach:1985uq,Deser:1986xr}. 

Despite the remarkable features mentioned above, two important disadvantages of the use of a vector-dilaton Weyl multiplet are  that: 
(i) the spectrum of the on-shell theory do not precisely match the one of minimal Poincar\'e supergravity as in fact it leads to an extra on-shell physical multiplet that includes a scalar (dilaton) field; 
(ii) it is not possible to describe gauged supergravity and then AdS supergravity.\footnote{It has been proposed in \cite{CO}, and successively described also in superspace in \cite{BKNT-M14}, how to gauge a system based on the vector-dilaton Weyl multiplet by appropriately deforming the constraint of the on-shell vector multiplet. However, so far, this construction has not been systematically studied as for gauged supergravities based on the standard Weyl multiplet, including curvature-squared invariants. Interestingly, the hyper-dilaton Weyl multiplet of \cite{Gold:2022bdk,Hutomo:2022hdi} has no apparent issues concerning gauging, at least for matter systems not including extra physical hypermultiplets.}. This second limitation has a clear impact if one is interested in using off-shell  supergravity in the study of AdS/CFT. Indeed, the authors of \cite{Bobev:2021qxx,Bobev:2022bjm,Cassani:2022lrk} employed a formulation of minimal gauged supergravity in 5D based on the standard Weyl multiplet, for which however they could  only use two of the three independent invariants, the ones of \cite{HOT,OP1,OP2}, explicitly known in terms of the component fields.
To this regard, it is worth explaining that, as first discussed in \cite{Baggio:2014hua}, see also \cite{Bobev:2020zov,Bobev:2021qxx,Bobev:2022bjm,Cassani:2022lrk}, the use of two invariants might suffice in 5D since a curvature dependent redefinition of the metric can reabsorb one of the three curvature-squared terms. It remains however a nontrivial open problem to prove this statement for whole locally supersymmetric invariants (e.g., including fermions) and to have clear control of the supersymmetry transformations under this redefinition. All three invariants might also play a role to construct general higher-derivative invariants beyond four derivatives.
It is also worth mentioning that the related recent analysis of \cite{Liu:2022sew} was based on the three independent curvature-squared invariants of \cite{HOT,BRS,OP1,OP2} defined using a vector-dilaton Weyl multiplet. However, it remains unclear to us whether the analysis of \cite{Liu:2022sew} might have some issues with supersymmetry, due to the constraints in defining the gauging (or equivalently the cosmological constant term) in a vector-dilaton Weyl formulation. 

Considering the potential subtleties in the recent studies in 
\cite{Bobev:2020zov,Bobev:2021qxx,Liu:2022sew,Bobev:2022bjm,Cassani:2022lrk} it is natural to look back at \cite{BKNT-M14} and elaborate on properties of the three independent curvature-squared invariants for minimal supergravity constructed in superspace. A fundamental property of these locally superconformal invariants is that they can all be constructed by using a standard Weyl multiplet, making straightforward their addition to the 5D minimal off-shell two-derivative gauged supergravity theory. In this paper we then start to report on new results based on these invariants. More specifically, we will present here detailed expressions of all the composite primary superfields associated to each invariant -- including a new expression for the $\log$ multiplet --  which can be readily used for component analyses. We will then describe the primary equations of motion in superspace that describe minimal 5D gauged supergravity deformed by an arbitrary combination of three curvature-squared invariants. Our results are defined in superspace, but they can be straightforwardly translated in components by using the analysis of \cite{BKNT-M14}. In particular, since all the expressions are fully covariant and described explicitly in terms of composites of descendants of the various multiplets, one could, for example, straightforwardly obtain the whole set of deformed supergravity equations of motion by the successive action of $Q$-supersymmetry. This will then be a new step towards several applications of the three locally superconformal invariants of \cite{BKNT-M14}. Moreover, we also introduce an alternative four-derivative invariant based on the linear multiplet compensator and the kinetic superfield of the vector multiplet compensator. This can be used to engineer a scalar curvature squared invariant also in alternative off-shell supergravities as for example the formulation based on the recently introduced 5D hyper-dilaton Weyl multiplet \cite{Hutomo:2022hdi}.

Our paper is organised as follows. In section \ref{Section--2} we describe the structure of 5D $\cN=1$ superconformal multiplets that will be used in this work. In section \ref{5Dactionsuperspace} we give the salient details of \cite{BKNT-M14} concerning the superspace construction of various locally superconformal invariants (including curvature-squared ones) which will play the role of action principles. One of our new results in section \ref{5Dactionsuperspace} includes the expression of a composite primary multiplet which defines the ``log multiplet'' curvature-squared invariant. Section \ref{5DEOMSuperspace} contains the main results of our paper: the superconformal primary equations of motion of all the curvature-squared terms for the minimal 5D gauged off-shell  supergravity based on the standard Weyl multiplet. An alternative construction of a scalar curvature square invariant is presented in section \ref{NewCurvature2}.
Our notation and conventions correspond to that of \cite{BKNT-M14} (see also \cite{Hutomo:2022hdi}, where a handful of typos from \cite{BKNT-M14} were fixed).

\section{Superconformal multiplets in 5D $\cN=1$ superspace}
\label{Section--2}
In this section we review several superconformal multiplets which will serve as building blocks for the various curvature-squared invariants presented in this work. After describing the standard Weyl multiplet of conformal supergravity in 5D $\cN=1$ conformal superspace, we move on to the discussion of the Abelian vector and off-shell linear multiplets. Here we make use of the approach and results given in \cite{BKNT-M14}. 
We refer the reader to \cite{KL,KT-M5D1,Howe5Dsugra,HL,KT-M_5D2,KT-M_5D3,KT-M08,K06} for other works on flat and curved superspace and off-shell multiplets in five dimensions.

\subsection{The standard Weyl multiplet}
\label{SWM-superspace}
The standard Weyl multiplet of 5D $\cN=1$ conformal supergravity \cite{Bergshoeff1} is associated with the gauging of the superconformal algebra $\rm F^2(4)$.  The multiplet contains $32+32$ physical components described by a set of independent gauge fields: the vielbein $e_\hm{}^\ha$,
the gravitino $\psi_\hm{}_\hal^i$,
the ${\rm SU(2)}_R$ gauge field $\phi_\hm{}^{ij}$, and a dilatation gauge field $b_\hm$. The other gauge fields associated with the remaining gauge symmetries: the spin connection  $\omega_\hm{}^{\ha\hb}$, the $S$-supersymmetry connection
$\phi_\hm{}_\hal^i$, and the special conformal connection
$\mathfrak{f}_\hm{}^\ha$ are composite fields, i.e., they are algebraically determined in terms of the other fields by imposing constraints on some of the
curvature tensors. The standard Weyl multiplet also comprises a set of covariant auxiliary fields: a real antisymmetric
tensor $w_{\ha\hb}$, a fermion $\chi_\hal^i$, and a real auxiliary scalar $D$. 

The 5D $\cN=1$ conformal superspace is parametrised by
local bosonic $(x^{\hm})$ and fermionic $(\theta_i)$ coordinates 
$z^{\hM} = (x^{\hm},\q^{{\mu}}_i)$, 
where $\hm = 0, 1, 2,3, 4$, ${\mu} = 1, \cdots, 4$, and $i = 1, 2$.
To perform the gauging of the
superconformal algebra, one introduces
covariant derivatives $ {\nabla}_{\hA} = (\nabla_{\ha} , \nabla_{\hal}^i)$ which have the form
\bsubeq
\bea\label{eq:covD}
\de_{\hA} 
= E_{\hA} - \o_{\hA}{}^{\underline{b}}X_{\underline{b}} 
&=& E_{\hA} - \hf \O_{\hA}{}^{\ha \hb} M_{\ha \hb} - \Phi_{\hA}{}^{ij} J_{ij} - B_{\hA} \mathbb{D} - \mathfrak{F}_{\hA}{}^{\hB} K_{\hB}~,
\\ &=& E_{\hA} - \hf \O_{\hA}{}^{\ha \hb} M_{\ha \hb} - \Phi_{\hA}{}^{ij} J_{ij} - B_{\hA} \mathbb{D} - \mathfrak{F}_{\hA}{}^{\a i} S_{\a i} - \mathfrak{F}_{\hA}{}^{a} K_{a} ~.~~~~
\eea
\esubeq
Here $E_{\hA} = E_{\hA}{}^{\hM} \partial_{\hM}$ is the inverse super-vielbein,
$M_{\ha \hb}$ are the Lorentz generators, $J_{ij}$ are generators of the
${\rm SU(2)}_R$ $R$-symmetry group,
$\mathbb D$ is the dilatation generator, and $K_{\hA} = (K_{\ha}, S_{\hal i})$ are the special superconformal
generators.
The super-vielbein one-form is $E^{\hA} =\rd z^{\hM} E_{\hM}{}^{\hA}$ with $E_{\hM}{}^{\hA} E_{\hA}{}^{\hN} =\d_{\hM}^{\hN}$ and
 $E_{\hA}{}^{\hM} E_{\hM}{}^{\hB}=\d_{\hA}^{\hB}$.
We associate with each generator $X_{\underline{a}} = (M_{\ha\hb},J_{ij},\bbD, S_{\hal i}, K_\ha)$ a connection super one-form 
$\omega^{\underline{a}} = (\O^{\ha\hb},\F^{ij},B,\mathfrak{F}^{\hal i},\mathfrak{F}^{\ha})= \rd z^\hM \omega_\hM{}^{\underline{a}} = E^{\hA} \o_{\hA}{}^{\underline{a}}$.

The algebra of covariant derivatives
\begin{align}
[ \nabla_\hA , \nabla_\hB \}
	&= -\mathscr{T}_{\hA\hB}{}^\hC \nabla_\hC
	- \frac{1}{2} {\mathscr{R}(M)}_{\hA\hB}{}^{\hc\hd} M_{\hc\hd}
	- {\mathscr{R}(J)}_{\hA\hB}{}^{kl} J_{kl}
	\non \\ & \quad
	- {\mathscr{R}(\mathbb{D})}_{\hA\hB} \mathbb D
	- {\mathscr{R}(S)}_{\hA\hB}{}^{\hga k} S_{\hga k}
	- {\mathscr{R}(K)}_{\hA\hB}{}^\hc K_\hc~,
	\label{nablanabla}
\end{align}
is constrained to be  expressed  in terms of a single primary superfield, the super-Weyl tensor $W_{\hal \hbe}$.\footnote{Here and in what follows, an antisymmetric rank-2 tensor $T_{a b} = -T_{b a}$ can equivalently be written as: $T_{a b} = (\Sigma_{a b}){}^{\a \b} T_{\a \b}$ and $T_{\hal \hbe} = 1/2 \,({\S}^{\ha \hb})_{\hal \hbe}T_{\ha \hb}$.} It has the following properties
\be
W_{\hal \hbe} = W_{\hbe \hal} \ , \quad
K_{\hA} W_{\hal \hbe} = 0 \ ,
 \quad \mathbb D W_{\hal \hbe} = W_{\hal \hbe} \ ,
\ee
and satisfies the Bianchi identity
\be 
\nabla_\hga^k W_{\hal \hbe} = \nabla_{(\hal}^k W_{\hbe \hga )} + \frac{2}{5} \eps_{\hga (\hal} \nabla^{\hde k} W_{\hbe ) \hde} \ . \label{WBI}
\ee
In  \eqref{nablanabla}
$ \mathscr{T}_{\hA \hB}{}^C$ is the torsion, and $ \mathscr{R}(M)_{\hA \hB}{}^{\hc \hd}$,
$ \mathscr{R}(J)_{\hA \hB}{}^{kl}$, $ \mathscr{R}(\mathbb D)_{\hA \hB}$, $ \mathscr{R}(S)_{\hA \hB}{}^{ {{\g}}k}$, and $ \mathscr{R}(K)_{\hA \hB}{}^{\hc}$
are the curvatures associated with Lorentz, ${\rm SU(2)}_R$,
dilatation, $S$-supersymmetry, and special conformal
boosts, respectively.

The full algebra of covariant derivatives \eqref{nablanabla} (including the explicit expressions for the torsion and curvature components in terms of the descendant superfields) are given in Refs. \cite{BKNT-M14} and \cite{Hutomo:2022hdi}. 
%%%
To make use of results of \cite{BKNT-M14} it is important to note that in this paper we make use of the ``traceless'' frame conventional constraints for the conformal superspace algebra employed in appendix C of \cite{BKNT-M14} as well as in \cite{Hutomo:2022hdi}. We also refer the reader to these papers for the description of how to reduce superspace results to standard component fields.

It is useful to introduce the dimension-3/2 superfields
\bsubeq \label{descendantsW-5d}
\begin{gather}
W_{\hal \hbe \hga}{}^k := \nabla_{(\hal}^k W_{\hbe \hga )} \ , \quad X_\hal^i := \frac{2}{5} \nabla^{\hbe i} W_{\hbe\hal} \ ,
\end{gather}
and the dimension-2 descendant superfields constructed from spinor covariant derivatives of $W_{\a \b}$:
\bea
W_{\hal \hbe \hga \hde} := \nabla_{(\hal}^k W_{\hbe \hga \hde) k} \ , \quad 
X_{\hal \hbe}{}^{ij} := \nabla_{(\hal}^{(i} X_{\hbe)}^{j)} 
\ , \quad
Y := \ri \nabla^{\hga k} X_{\hga k} \ .
\eea
\esubeq
It can be checked that only the superfields \eqref{descendantsW-5d} and their vector derivatives appear upon taking successive spinor derivatives of $W_{\a \b}$. The following relations define the tower of covariant fields in the standard Weyl  multiplet and are particularly useful for analysing the structure of curvature-squared invariants:
\bsubeq \label{eq:Wdervs}
\bea
\nabla_{\hga}^k W_{\hal\hbe} 
&=& W_{\hal\hbe\hga}{}^k + \eps_{\hga (\hal} X_{\hbe)}^k \ ,  \label{Wdervs-a}
\\
\nabla^{i}_{\a}{X^{j}_{{\beta}}} 
&=& 
X_{{\alpha} {\beta}}\,^{i j}
+\frac{\ri}{8} \eps^{i j} \eps_{{\alpha} {\beta}} Y
-\frac{3\ri}{2} \eps^{i j} (\Gamma^{{a}})_{{\alpha}}{}^{ \rho} {\nabla}_{{a}}{W_{{\beta} {\rho}}}
-2\ri \eps^{i j} W_{{\alpha}}{}^{{\rho}} W_{{\beta} {\rho}}
\non\\
&&
+\frac{\ri}{2} \eps^{i j} \eps_{{\alpha} {\beta}} W^{{\gamma} {\delta}} W_{{\gamma} {\delta}}
-\frac{\ri}{2}\eps^{i j} (\Gamma^{{a}})_{{\beta}}{}^{{\rho}} {\nabla}_{{a}}{W_{{\alpha} {\rho}}}
~,
\\
\nabla^{i}_{{\alpha}}{W_{{\beta} {\gamma} {\lambda}}{}^{j}}
&=& -  \frac{1}{2} \eps^{i j} \Big( W_{{\alpha} {\beta} {\gamma} {\lambda}} + 3 \ri  (\Gamma_{{a}})_{{\alpha} ({\beta}}  {\nabla}^{{a}}{W_{{\gamma} {\lambda})}} + 3 \ri \eps_{{\alpha} ({\beta}} (\Gamma_{{a}})_{{\gamma}}{}^{{\tau}}   {\nabla}^{{a}}{W_{{\lambda}) {\tau}}}\Big) \non \\&&- \frac{3}{2}\eps_{{\alpha} ({\beta}} X_{{\gamma} {\lambda})}\,^{i j} \ , 
\\ 
\nabla^{i}_{{\alpha}}{W_{{\beta} {\gamma} {\lambda} {\rho}}} 
&=& - 4 \ri (\Gamma_{{a}})_{{\alpha} ({\beta}} {\nabla}^{{a}}{W_{{\gamma} {\lambda} {\rho})}\,^{i}}- 6 \ri  W_{{\alpha} ({\beta} {\gamma}}\,^{i} W_{{\lambda} {\rho})}+6 \ri W_{{\alpha} ({\beta}} W_{{\gamma} {\lambda} {\rho})}\,^{i}\non \\&& +6 \ri \eps_{{\alpha} ({\beta}} \Big( W_{{\gamma} {\lambda}} X^{i}_{{\rho})}-2 (\Gamma_{{a}})_{{\gamma}}{}^{{\tau}}   {\nabla}^{{a}}{W_{{\lambda} {\rho}) {\tau}}\,^{i}}-  W_{{\gamma}}{}^{{\tau}} W_{{\lambda} {\rho} ){\tau}}\,^{i}\Big) \ , \\
\nabla^{i}_{{\alpha}}{X_{{\beta} {\gamma}}\,^{j k}} 
&=& \ri  \eps^{i (j} \Big( -3   W_{({\beta}}{}^ {\lambda} W_{ {\gamma}) {\alpha}\lambda}\,^{ k)} -  \eps_{{\alpha} ({\beta}}  W^{{\rho} {\tau}} W_{{\gamma}) {\rho} {\tau}}\,^{k)}  -  W_{{\alpha} {\lambda}} W_{{\beta} {\gamma} }\,^{{\lambda}k)}  - \frac{3}{2} W_{{\beta} {\gamma}} X^{k)}_{{\alpha}}  
\non\\&& \qquad ~
+\frac{1}{2} W_{{\alpha} ({\beta}} X^{k)}_{{\gamma})} +\frac{3}{2} \eps_{{\alpha} ({\beta}}  W_{{\gamma}) {\lambda}} X^{k) {\lambda}}  + 2 (\Gamma^{{a}})_{{\alpha}}{}^{{\rho}}  {\nabla}_{{a}}{W_{{\beta} {\gamma} {\rho}}\,^{k)}} 
\non\\&& \qquad  ~
+ 2  (\Gamma^{{a}})_{({\beta} }{}^{{\rho}}   {\nabla}_{{a}}{W_{ {\gamma}) {\alpha} {\rho}}\,^{k)}}  -  (\Gamma^{{a}})_{{\alpha} ({\beta}}  {\nabla}_{{a}}{X^{k)}_{{\gamma})}}
+  \eps_{{\alpha} ({\beta}} (\Gamma^{{a}})_{{\gamma}) {\lambda}}  {\nabla}_{{a}}{X^{k) {\lambda}}} \Big) 
~,~~~~~~ \\
\nabla^{i}_{{\alpha}} {Y} 
&=& 8(\Gamma^{{a}})_{{\alpha}}{}^{{\beta}}  {\nabla}_{{a}}{X^{i}_{ {\beta}}} + 8 W_{{\alpha}}{}^{{\beta}} X^{i}_{{\beta}}
~.
\eea
\esubeq
Due to \eqref{WBI}, the $X_{\a\b}{}^{ij}$ and $W_{\a\b\g\d}$ dimension-2 superfields of the standard Weyl multiplet obey the following Bianchi identities:
\bsubeq
\bea
  \nabla_{(\a}{}^{\g} X_{\b)\g}{}^{ij} &=& -\frac{1}{2} X^{\g (i} W_{\a \b \g}{}^{j)}
  ~,\\
   \nabla_{(\a}{}^{\l} W_{\b \g \tau)\l} &=& 3 \ri  \nabla_{(\a}{}^{\l} \Big(W_{\b \g} W_{\tau)\l}\Big)
   ~.
 \eea
\esubeq
The independent descendant superfields of $W_{\a\b}$ are all annihilated by $K_{\ha}$. However, under $S$-supersymmetry, they transform as follows:
\bsubeq 
\bea 
S_{\hal i} W_{\hbe\hga\hde}{}^j 
&=& 
6 \d^j_i \eps_{\hal (\hbe} W_{\hga \hde)} \ , \qquad
S_{\hal i} X_\hbe^j = 4 \d_i^j W_{\hal\hbe}~, 
\\
S_{\hal i} W_{\hbe\hga\hde\hrh} &=& 24 \eps_{\hal (\hbe} W_{\hga\hde \hrh)}{}_i \ , \qquad
S_{\hal i} Y = 8 \ri X_{\hal i} 
~, 
\\
S_{\hal i} X_{\hbe\hga}{}^{jk} &=& - 4 \d_i^{(j} W_{\hal\hbe\hga}{}^{k)} + 4 \d_i^{(j} \eps_{\hal (\hbe} X_{\hga)}^{k)} \ .\label{S-on-X_Y-a}
\eea 
\esubeq

The conformal supergravity gauge group $\cG$ is generated by
{\it covariant general coordinate transformations},
$\delta_{\rm cgct}$, associated with a local superdiffeomorphism parameter $\xi^{\hA}$ and
{\it standard superconformal transformations},
$\delta_{\cH}$, associated with the local superfield parameters:
the dilatation $\s$, Lorentz $\L^{\ha \hb}=-\L^{\hb \ha}$,  ${\rm SU(2)}_R$ $\L^{ij}=\L^{ji}$,
 and special conformal transformations $\L^{\hA}=(\eta^{\hal i},\L^{\ha}_{K})$.
The covariant derivatives transform as
\bsubeq
\label{sugra-group}
\bea
\d_\cG \nabla_{\hA} &=& [\cK , \nabla_{\hA}] \ ,
\label{TransCD}
\eea
where 
\bea
\cK = \xi^{\hC}  {\nabla}_{\hC} + \hf  {\L}^{\ha \hb} M_{\ha \hb} +  {\L}^{ij} J_{ij} +  \s \mathbb D +  {\L}^{\hA} K_{\hA} ~.
\eea
\esubeq
A covariant (or tensor) superfield $U$ transforms as
\be
\d_{\cG} U =
(\d_{\rm cgct}
+\d_{\cH}) U =
 \cK U
 ~. \label{trans-primary}
\ee
The superfield $U$ is a \emph{superconformal primary} of dimension $\D$ if $K_{\hA} U = 0$ (it suffices to require that $S_{\a i} U = 0$) and $\mathbb D U = \D U$.

\subsection{The Abelian vector multiplet}
\label{vector-superspace}

In conformal superspace \cite{BKNT-M14}, a 5D $\cN=1$ Abelian vector multiplet \cite{HL, Zupnik:1999iy} is described by a real primary superfield $W$ of dimension 1,
\bsubeq\label{vector-defs}
\bea
(W)^* = W~, \qquad K_A W=0~, \qquad \bbD W= W~.
\eea
The superfield $W$ obeys the Bianchi identity
\bea 
\nabla_{{\a}}^{(i} \nabla_{{\b}}^{j)} W 
= \frac{1}{4} \eps_{{\a} {\b}} \nabla^{{\g} (i} \nabla_{{\g}}^{j)} 
W
~.
\label{vector-Bianchi}
\eea
\esubeq

Let us introduce the following descendants constructed from spinor derivatives of $W$:
 \bsubeq \label{components-vect}
\begin{align}
\l_\hal^i := - \ri\nabla_\hal^i W \ , \qquad
X^{ij} := \frac{\ri}{4} \nabla^{\hal (i} \nabla_\hal^{j)} W 
= - \frac{1}{4} \nabla^{\hal (i}  \l_\hal^{j)}~.
\end{align}
These superfields, along with
\be 
F_{\hal\hbe} := - \frac{\ri}{4} \nabla^k_{(\hal} \nabla_{\hbe) k}  W - W_{\a \b} W
= \frac{1}{4} \nabla_{(\hal}^k \l_{\hbe) k} 
- W_{\a \b} W
~,
\ee
\esubeq
satisfy the following identities:
\bsubeq \label{VMIdentities}
\begin{align}
\nabla_\hal^i \l_\hbe^j 
&= - 2 \eps^{ij} \big(F_{\hal \hbe} + W_{\hal\hbe}  W\big) 
- \eps_{\hal\hbe} X^{ij} - \eps^{ij} \nabla_{\hal\hbe} W \ , \\
\nabla_\hal^i F_{\hbe\hga} 
&=
 - \ri \nabla_{\hal (\hbe} \l_{\hga)}^i 
 - \ri \eps_{\hal (\hbe} \nabla_{\hga )}{}^\hde \l_\hde^i 
- \frac{3\ri}{2} W_{\hbe\hga} \l_\hal^i - W_{\hal\hbe \hga}{}^i  W \non\\
&~~~+ \frac{\ri}{2} W_{\a (\b} \l_{\g)}^i
-\frac{3 \ri}{2} \ve_{\a (\b} W_{\g)}{}^{\d} \l_{\d}^i\ , \\
\nabla_\hal^i X^{jk} &= 2 \ri \eps^{i (j} 
\Big(
\nabla_\hal{}^\hbe \l_\hbe^{k)} 
- \hf W_{\hal\hbe} \l^{\hbe k)} 
+ \frac{3\ri}{4} X_\hal^{k)} W
\Big) 
\ .
\end{align}
\esubeq
We also note that $F_{\a \b} = \hf (\S^{ab})_{\a \b} F_{ab}$. Due to \eqref{vector-Bianchi}, a dimension-2 superfield of a vector multiplet in the traceless frame obeys the following Bianchi identity:
\begin{align}
         \nabla_{(\a}{}^{\g} F_{\b)\g} = \frac{1}{2} \l^{\g k} W_{\a \b \g k}
         ~.
\end{align}

The actions of the $S$-supersymmetry generator on the descendants are given by
\begin{align}
S_\hal^i \l_\hbe^j &= - 2 \ri \eps_{\hal\hbe} \eps^{ij}  W \ , \qquad
S_\hal^i F_{\hbe\hga} = 4 \eps_{\hal (\hbe}  \l_{\hga)}^i \ , \qquad
S_\hal^i X^{jk} = - 2 \eps^{i (j}  \l_\hal^{k)} \ ,
\label{S-on-W}
\end{align}
while all the fields are annihilated by the $K_a$ generators.

In 5D $\cN=1$ conformal superspace, there exists a prepotential formulation for the Abelian vector multiplet, which,  was developed in \cite{BKNT-M14}, see also \cite{K06,KL,KT-M08,KT-M_5D3} for earlier related analysis in other superspaces. The authors of \cite{BKNT-M14} introduced a real primary superfield $V_{ij}$ of dimension $-2$, $\mathbb{D} V_{ij} = -2 V_{ij}$. It was also shown that $V_{ij}$ transforms as an isovector under ${\rm SU}(2)_R$ transformations and is the 5D analogue of Mezincescu's prepotential \cite{Mezincescu, HST, BK11} for the 4D $\cN=2$ Abelian vector multiplet. This then allows us to represent the field strength $W$ as
\bea
W= -\frac{3\ri}{40} \de_{ij} \D^{ijkl} V_{kl}~,
\label{mezincescu}
\eea
where we have defined the operators
 \bsubeq
 \bea
 \D^{ijkl} &:=& 
 -\frac{1}{96} \ve^{\a \b \g \d} \de_{\a}^{(i} \de_{\b}^j \de_{\g}^k \de_{\d}^{l)} = -\frac{1}{32} \de^{(ij} \de^{kl)} = \D^{(ijkl)}~, \\
 \de^{ij} &:=& \de^{\a (i} \de_{\a}^{j)}~.
 \eea
 \esubeq
It should be noted that $V_{ij}$ in \eqref{mezincescu} is defined modulo gauge transformations of the form 
\bea
\d V_{kl} = \de_{\a}^p \L^{\a}{}_{klp}~, \qquad \L^{\a}{}_{klp} = \L^{\a}{}_{(klp)}~,
\label{gauge-vector}
\eea
with the gauge parameter $\L^{\a}{}_{klp}$ being a primary superfield,
\bea
S_{\a}^i \L^{\b}{}_{jkl}=0~, \qquad  \mathbb{D} \L^{\b}{}_{jkl} = -\frac{5}{2} \L^{\b}{}_{jkl}~.
\eea

\subsection{The linear multiplet}

The linear multiplet \cite{FS2,deWit:1980gt,deWit:1980lyi,deWit:1983xhu,N=2tensor,Siegel:1978yi,Siegel80,SSW,deWit:1982na,KLR,LR3}, 
 or $\cO(2)$ multiplet, can be described in terms of the primary superfield $G^{ij}= G^{ji}$, which is characterised by the properties
\bsubeq\label{linear-constr}
\bea
\de_{\a}^{(i} G^{jk)} &=& 0~, 
\label{O(2)constraints}\\
K_A G^{ij} &=& 0~, \qquad \mathbb{D} G^{ij} = 3 G^{ij}~.
\eea
\esubeq
We assume $G^{ij}$ to be real, $(G^{ij})^{*}= \ve_{ik} \ve_{jl} G^{kl}$.

The component structure of $G^{ij}$ is characterised by the following tower of identities:
\bsubeq \label{O2spinorderivs}
\begin{align}
\nabla_\hal^i G^{jk} &= 2 \eps^{i(j} \varphi_\hal^{k)} \ , \\
\nabla_\hal^i \varphi_\hbe^j &= - \frac{\ri}{2} \eps^{ij} \eps_{\hal\hbe} F + \frac{\ri}{2} \eps^{ij} \cH_{\hal\hbe} + \ri \nabla_{\hal\hbe} G^{ij} \ , \\
\nabla_\hal^i F &= - 2 \nabla_\hal{}^\hbe \varphi_\hbe^i - 3 W_{\hal\hbe} \varphi^{\hbe i} - \frac{3}{2} X_{\hal j} G^{ij} \ , \\
\nabla_\hal^i \cH_{\ha} &= 4 (\S_{\ha\hb})_\hal{}^\hbe \nabla^\hb \varphi_\hbe^i - \frac{3}{2} (\G_\ha)_\hal{}^\hbe W_{\hbe \hga} \varphi^{\hga i}
-\frac{1}{2} (\G_\ha)_\hga{}^\hbe W_{\hbe \hal} \varphi^{\hga i} \ ,
\end{align}
\esubeq
where we have defined the independent descendants superfields
\bsubeq \label{O2superfieldComps}
\begin{align}
\varphi_\hal^i &:= \frac{1}{3} \nabla_{\hal j} G^{ij} \ , \\
F &:= \frac{\ri}{12} \nabla^{\hga i} \nabla_\hga^j G_{ij} = - \frac{\ri}{4} \nabla^{\hga k} \varphi_{\hga k} \ ,\\
\cH_{abcd} &:= \frac{\ri}{12} \eps_{abcde} (\G^{e})^{\a \b} \nabla_{\a}^i \nabla_{\b}^j G_{ij} 
\equiv \eps_{\ha\hb\hc\hd\he} \cH^\he \ .
\end{align}
\esubeq
Here $\cH^{a}$ obeys the differential condition 
\be 
\nabla_\ha \cH^\ha = 0~, \qquad {\cH}^{\ha}:= -\frac{1}{4!}\ve^{\ha \hb \hc \hd \he} \cH_{\hb \hc \hd \he}~.
\ee
The descendants \eqref{O2superfieldComps} are all annihilated by $K_a$. Under the action of $S$-supersymmetry, they transform as follows:
\begin{align} S_\hal^i \varphi_\hbe^j = - 6 \eps_{\hal \hbe} G^{ij} \ , ~~~~~~
S_\hal^i F = 6 \ri \varphi_\hal^i \ , ~~~~~~
S_\hal^i \cH_\hb &= - 8 \ri (\G_\hb)_\hal{}^\hbe \varphi_\hbe^i \ .
\label{O2-S-actions}
\end{align}
We refer the reader to \cite{BKNT-M14} for a superform description of the linear multiplet.

As described in \cite{BKNT-M14}, the linear multiplet constraints \eqref{linear-constr} may be solved in terms of  
 an arbitrary primary real dimensionless scalar prepotential $\O$, 
\bea
S_\hal^i\O=0~,~~~~~~
\bbD\O=0
~,
\eea
and the solution is 
\bea
G^{ij}
&=&
-\frac{3\ri}{40}\D^{ijkl}\de_{kl}\O ~.
\label{def-G-0-b}
\eea
A crucial property of $G^{ij}$ defined by 
\eqref{def-G-0-b} is that it is invariant under  gauge transformations 
of $\O$
of the form 
\bea
\d\Omega=-\frac{\ri}{2}(\G^\ha)^{\hal\hbe}\de_\hal^i \de_\hbe^jB_\ha{}_{ij}
~,
\label{gauge-O2-0}
\eea
where the gauge parameter is assumed to have the properties
\bea
B_{\ha}{}^{ij}=B_{\ha}{}^{ji}
~,~~~~~~
S_{\hal}^{i} B_\ha{}^{jk}=0
~,~~~~~~
\bbD B_\ha{}^{ij}=-B_\ha{}^{ij}
~,
\label{gauge-inv-O2-1}
\eea
and is otherwise arbitrary.

To conclude this section we introduce another result that will be used in the rest of the paper. Given a system of $n$ Abelian vector multiplets $W^I$, with $I=1,2, \dots n$,  all satisfying \eqref{vector-defs}, we can construct the following composite linear multiplet and its descendants \cite{BKNT-M14}:
\bsubeq \label{O2composite-N}
\bea
H^{ij}
&=& C_{JK} \Big\{ \,2 W^J X^{ij\, K}
- \ri \l^{\a J\,(i } \l_{\a}^{j) K}\Big\}
~,\\
%%%%%%%%%%%%%%%%%%%%%%%
\vf_{\a\, }^i &=& C_{JK} \bigg\{\,
\ri X^{ij\, J} \l_{\a j}^K 
-2 \ri F_{\a \b}^{J} \l^{\b i K}
- \frac{3}{2} X_{\a}^i W^J W^K
-2 \ri W^J \de_{\a \b} \l^{\b i K}\non\\
&&~~~~~~~~
- \ri (\de_{\a \b} W^J) \l^{\b i K}
- 3 \ri W_{\a \b} W^J \l^{\b i K}
\bigg\}~, \\
%%%%%%%%%%%%%%%%%%%%%%%%%%%%%%%%%%%%
F &=& C_{JK} \bigg\{\,
X^{ij J}X_{ij}^{K}
- F^{ab J} F_{ab}^{K}
+ 4 W^J \Box W^K
+ 2 (\de^a W^J) \de_{a} W^K \non\\
&&~~~~~~~~
+ 2 \ri (\de_{\a}{}^{\b} \l_{\b}^{iJ}) \l^{\a K}_{i}{} -6 W^{ab} F_{ab}^J W^K 
-\frac{39}{8} W^{ab} W_{ab} W^J W^K
 \non\\
&&~~~~~~~~
+ \frac{3}{8} Y W^J W^K
+ 6 X^{\a i}\l_{\a i}^J W^K
-3 \ri W_{\a \b} \l^{\a i J} \l^{\b K}_{i}
\bigg\}
~, \\
%%%%%%%%%%%%%%%%%%%%%%%%%%%%%%%%%%%%%%
\cH_{a} &=& C_{JK} \bigg\{
-\hf \ve_{abcde} F^{bc\, J} F^{de \,K}
+ 4 \de^{b} \Big( W^J F_{ba}^K + \frac{3}{2} W_{ba} W^J W^K\Big)\non\\
&&~~~~~~~~~
+ 2 \ri (\S_{ba})^{\a \b} \de^{b} (\l_{\a}^{iJ} \l_{\b i}^K)
\bigg\}
~,
\eea
\esubeq
where $\Box := \de^{a} \de_{a}$ and $C_{JK} = C_{(JK)}$ is a constant symmetric in $J$ and $K$.
Equation \eqref{O2composite-N} is the superspace analogue of the composite linear multiplet constructed in \cite{Bergshoeff1}.

\section{Superconformal actions} 
\label{5Dactionsuperspace}

In this section, we review a main action principle that was used in \cite{BKNT-M14} to construct various locally superconformally invariants (including curvature-squared ones) in superspace. A simple way to define it is based on a full superspace integral
\begin{align}
S[\cL] = \int\rd^{5|8}z\,  E\, \cL~,\qquad\rd^{5|8}z:=\rd^5x\,\rd^8\q~, \qquad
E := {\rm Ber}(E_\hM{}^\hA)~,
\end{align}
where the Lagrangian $\cL$ is a conformal primary superfield of dimension $+1$, $\mathbb{D} \cL= \cL$. This invariant can be proven to be locally superconformal invariant, that is, invariant under the supergravity gauge transformations \eqref{sugra-group}.

\subsection{BF action}

The action involving the product of a linear multiplet with an Abelian vector multiplet is referred to as the BF action. Analogous to the component superconformal tensor calculus, this plays a fundamental role in the construction of general supergravity-matter couplings, see \cite{Ohashi1,Ohashi2,Ohashi3,Ohashi4,Bergshoeff1,Bergshoeff2,Bergshoeff3} for the 5D case, and it was a main building block for the  invariants introduced in \cite{BKNT-M14} that we focus on.
In superspace the BF action may be described by 
\bsubeq\label{BF-action-0}
\bea
S_{\rm{BF}}
=
\int\rd^{5|8}z\, E\, 
 \O W
=
 \int \rd^{5|8}z\, E\, 
 G^{ij} V_{ij}
 ~.
  \label{SGV}
\eea
As implied by the equation above, the BF action can be written in different ways, see \cite{BKNT-M14} for even more variants. In the first form in \eqref{SGV}, it involves the field strength of the vector multiplet, 
$W$, and the prepotential of the linear multiplet, $\O$.  
By using \eqref{def-G-0-b}, \eqref{mezincescu}, and then integrating by parts, one may obtain the equivalent form of the BF action involving Mezincescu's prepotential $V_{ij}$ and the field strength $G^{ij}$ described by the right-hand side of \eqref{SGV}. 
One may also prove that the  functionals $\int \rd^{5|8}z\, E\, 
 \O W$ and 
 $\int \rd^{5|8}z\, E\, 
 G^{ij} V_{ij}$
are, respectively, invariant under the gauge transformations \eqref{gauge-O2-0} and \eqref{gauge-vector}, thanks to the defining differential constraints satisfied by $W$ and $G^{ij}$, eqs.~\eqref{vector-Bianchi} and \eqref{O(2)constraints}.

In components, and in our notation, the BF action takes the form \cite{BKNT-M14} 
\begin{equation}
\begin{aligned}
    S_{\rm BF} = - \int \rd^5 x & \, e \Big( \,v_{{a}} \cH^{{a}} + W F + X_{i j} G^{i j} + 2 \l^{\a k} \vf_{\a k} \\
    &- \psi_{{a }}{}^{\a}_i(\Gamma^{{a}})_{\a}{}^{\b} \vf_{\b}^i  W - \ri \psi_{{a}}{}^{\a}_{i} (\Gamma^{{a}})_{\a}{}^{\b} \l_{\b j} G^{i j} +  \ri \psi_{{a}}{}^{\a}_{i} (\Sigma^{{a} {b}}){}_{\a}{}^{\b} \psi_{{b} \b j} W G^{i j} \Big)~. \label{BF-Scomp}
\end{aligned}
\end{equation}
\esubeq 
In \eqref{BF-Scomp}, we have defined the usual component projection to $\q = 0$, i.e., $U(z) | := U(z) |_{\q=0}$. We associate the same symbol for the covariant component fields and the corresponding superfields, when the interpretation is clear from the context. 
Here $v_{\hm} := V_m|$ denotes a real Abelian gauge connection. Its real field strength is $f_{mn} :=  F_{mn}| = 2 \pa_{[m} v_{n]}$.  Note that the  field strength $f_{mn}$ may be expressed in terms of the bar-projected, covariant field strength $F_{ab}:= F_{ab} \loco$ via the relation
\bea
F_{ab} = f_{ab} + \ri (\G_{[a})_{\a}{}^{\b} \psi_{b]}{}^{\a}_{k} \l_{\b}^k  
+ \frac{\ri}{2} W\, \psi_{[a}{}^{\g}_{k} \psi_{b]}{}^{k}_{\g}~, \qquad f_{ab}:= e_{a}{}^{m} e_{b}{}^{n} f_{mn}~.
\eea
When projected to components, the lowest component of the covariant superfield $\cH_a$ satisfies the constraint $\de^{\ha} {\cH}_{\ha} = 0$~, where $\cH_a := \cH_a \loco$. 
It holds that
\bea
\cH^a = h^a 
+ 2 (\S^{ab})_\a{}^\b\psi_{b}{}^\a_i\vf_\b^i
- \frac{\ri}{2} \ve^{abcde} (\S_{bc})_{\a\b}\psi_{d}{}^\a_{i}\psi_{e}{}^\b_{j} G^{ij} ~.
\eea
The constraint $\de^{a} \cH_a=0$ implies the existence of a gauge three-form potential, $b_{\hm \hn \hp}$, and its exterior derivative $h_{\hm \hn \hp \hq}:= 4 \pa_{[\hm}b_{\hn \hp \hq]}$.
See \cite{BKNT-M14} and \cite{Hutomo:2022hdi} for more details.

\subsection{Vector multiplet compensator} 
\label{vector_compensator}

The two-derivative invariant for the vector multiplet compensator can be constructed using the above BF action principle \eqref{SGV} but with the linear multiplet being a composite superfield. 
We denote by $H_{\rm{VM}}^{ij}$ the composite linear multiplet \eqref{O2composite-N}, which is built out of a single Abelian vector multiplet:
\bea 
H^{ij}_{\rm{VM}} &=& \ri (\de^{\a(i} W) \de_{\a}^{j)} W + \frac{\ri}{2} W \de^{\a(i} \de_{\a}^{j)} W \non\\
&=& -\ri {\lambda}^{\a i} {\l}_{\a}^j +2 {W} {X}^{ij}~.\label{HijVM}
\eea
One can check that $H_{\rm{VM}}^{ij}$  is a dimension-3 primary superfield, $S_{\a}^k H^{ij}_{\rm{VM}}= 0$. Thanks to the Bianchi identity \eqref{vector-Bianchi} obeyed by the field strength $W$, the composite superfield $H_{\rm{VM}}^{ij}$ satisfies the analyticity constraint
\bea
\de^{(i}_{\a} H^{jk)}_{\rm{VM}}=0~.
\eea

The vector multiplet action may then be rewritten as an integral over the full superspace,
\bea
S_{\rm{VM}}&=&
\frac{1}{4} \int\rd^{5|8}z\, E\,   V_{ij} { H }_{\rm VM}^{ij} ~.~~~
\label{rep8.3}
\eea
It is also possible to write the action as
\bea
S_{\rm{VM}}= 
\frac{1}{4}
 \int \rd^{5|8}z\, E\, 
 {\bm \O}_{\rm VM} W~,
 \label{rep8.4}
\eea
where we have introduced the primary superfield $ {\bm \O}_{\rm VM}$ defined by \cite{BKNT-M14}:
\bea
 {\bm \O}_{\rm VM} = 
\frac{\ri}{4}\Big(
 W\de^{ij} V_{ij}
-2  (\de^{\hal i}V_{ij})\de_\hal^{j}W
-2 V_{ij}\de^{ij} W \Big)
~.
\eea
This is a prepotential for $H_{\rm{VM}}^{ij}$ in the sense of \eqref{def-G-0-b}.

The representations \eqref{rep8.3} and \eqref{rep8.4} allow us to compute the variation of 
$S_{\rm{VM}}$ with respect to the Mezincescu's prepotential, 
\bea
\d S_{\rm{VM}}
&=&
\frac{3}{4}\int \rd^{5|8}z\, E\, \d V_{ij}
{ H }_{\rm VM}^{ij} ~.
\label{var8.6b}
\eea
Note that the above variation vanishes when $\d V_{ij}$  is a gauge transformation \eqref{gauge-vector}. This implies that
\bea
\int \rd^{5|8}z\, E\, \L^{\a}{}_{ijk} \de_{\a}^{(k} H_{\rm VM}^{ij)} = 0~,
\eea
that is, $\de_{\a}^{(i} H_{\rm VM}^{jk)} = 0$.
This result is true for  any dynamical system involving an Abelian vector multiplet \cite{BKNT-M14}. The variation with respect to the prepotential $V_{ij}$ couples to a composite linear multiplet which depends on the specific form of the associated action principle -- let us call this, in general, $\mathbf{H}^{ij}$ which satisfies by construction the constraints \eqref{linear-constr}.
The equation of motion (EOM) for a vector multiplet is then $\mathbf{H}^{ij}=0$.
In the case of eq.~\eqref{rep8.3}, the EOM for the vector multiplet compensator is $H^{ij}_{\rm VM}=0$.

The superspace action $S_{\rm VM}$ can be reduced to components. The bosonic part of the component action reads \cite{BKNT-M14}
\bea 
S_{\rm VM} &=& \int \rd^{5} x\, e\, \bigg\{
- \frac{1}{8} W^3 {\cR}
+ \frac{3}{2} W ({\cD}^{\ha} W) {\cD}_{\ha} W
- \frac{3}{4} W X^{ij} X_{ij}
+\frac{1}{8} \ve_{\ha \hb \hc \hd \he}v^{\ha} f^{\hb \hc} f^{\hd \he} 
\non\\
&&~~~~~~~~~~~~~
+ \frac{3}{4} W f^{\ha \hb} f_{\ha \hb}
+ \frac{9}{4} W^2 W^{\ha \hb} f_{\ha \hb}
+ \frac{39}{32} W^3 W^{\ha \hb} W_{\ha \hb} - \frac{3}{32} W^3 Y \bigg\}~, 
\label{bosonic-swm}
\eea
where $\cR$ denotes the scalar curvature. In the above, we have introduced the  spin, dilatation, and
${\rm SU}(2)_R$ covariant derivative ${\cD}_\ha$ 
\begin{align}
\cD_{\ha} &= e_{\ha}{}^{\hm} \cD_{\hm} = e_{a}{}^{m} \Big(\pa_\hm
	- \frac{1}{2} \omega_\hm{}^{\hb \hc} M_{\hb \hc}
	- b_\hm \mathbb D
	- \phi_\hm{}^{ij} J_{ij} \Big)~.
\end{align}
The action is two-derivative and, upon gauge fixing dilatation by imposing $W=1$, the first term gives a scalar curvature term, $\cR$.
The gauge fixing $W=1$ can be achieved by requiring $W\ne 0$ meaning that the vector multiplet is a conformal compensator.

%%%%%%%%%%%%%%%%%%%%%%%%%%%%%%%%%%%%%%%%%%%%%%%%%%%%%%%%%%%%%
\subsection{Linear multiplet compensator} 
\label{linear_compensator}

The action for the linear multiplet compensator can also be constructed using the BF action principle \eqref{SGV}. In this case, the dynamical part of the action is described by a vector multiplet built out of the linear multiplet. 
We denote by $\mathbf{W}$ the composite vector multiplet field strength:
\bea
\mathbf{W} := 
 \frac{\ri}{16} G \nabla^{\hal i} \nabla_\hal^j \Big(\frac{G_{ij}}{G^2}\Big) = \frac{1}{4}F G^{-1} - \frac{\ri}{8} G_{i j} \varphi^{i \a} \varphi^{j}_{\a} G^{-3}~,
 \label{comp-W}
\eea
with
\bea
G := \sqrt{\hf G^{ij} G_{ij}}
\eea
being nowhere vanishing, $G \neq 0$. At the component level, the vector multiplet \eqref{comp-W} was first derived by Zucker
\cite{Zucker:2000ks} as a 5D analogue of the improved 4D $\cN=2$ tensor multiplet \cite{deWit:1982na}. 
The field strength $\mathbf{W}$ obeys 
the constraints \eqref{vector-defs}.

The action for the linear multiplet compensator may then be rewritten as 
\bea \label{S_boldW}
S_{\rm L} =  \int \rd^{5|8}z\, E\,  \O \mathbf{W} ~.
\eea
Varying the prepotential $\O$ leads to
\bea
\d S_{\rm L}
= 
 \int \rd^{5|8}z\, E\, \d \O \,\mathbf{W} ~.
\label{rep8.11}
\eea
Similar to what we discussed for the vector multiplet case, the previous form holds for the first-order variation of a matter system which includes a linear multiplet with respect to its prepotential $\O$. In particular, the variation must vanish if $\d \O$ is the gauge transformation \eqref{gauge-O2-0}. This holds if $\mathbf{W}$ obeys the Bianchi identity \eqref{vector-Bianchi}. 
In general, any dynamical system involving a linear multiplet then possesses a composite vector multiplet $\mathbf{W}$. The EOM for the linear multiplet is $\mathbf{W}=0$ and for the specific case of the linear multiplet action of eq.~\eqref{S_boldW} this is given by $\mathbf{W}$ defined in \eqref{comp-W}.

The bosonic part of $S_{\rm L}$ is given by \cite{BKNT-M14}
\bea
\label{eq:TensorComp}
S_{\rm L} 
&=& 
\int \rd^{5}x\, e\, 
\bigg\{ 
- \frac{3}{8} G\cR 
+\frac{3}{32} GY 
 - \frac{1}{8G} F^2
- \frac{3}{32} W^{\ha\hb} W_{\ha\hb} G
+\frac{1}{4} G^{-1}( \cD_\ha G^{ij} )\cD^\ha G_{ij}
\eol &&
- \frac{1}{2} G^{-1} \cH^\ha \cH_\ha
 + \frac{1}{12} \ve^{\ha\hb\hc\hd\he} b_{\hc\hd\he} \Big(
\frac{1}{2} G^{-3} (\cD_\ha G_{ik}) (\cD_\hb G_j{}^k) G^{ij}
+ G^{-1} R(J)_{\ha\hb}{}^{ij} G_{ij}
\Big)
\bigg\}~.~~~~~~~~~
\eea
%%%%%%%%%%%%%%%%%%%%%%%%%%%%%
The action is two-derivative, and, with $G=1$, the first term gives an $\cR$ term.

For later use it is useful to provide the explicit expressions of the composite descendant superfields of $\mathbf{W}$. These are given by
\bsubeq \label{desc-W-R2}
\bea
\bm{\l}_{\a}^i &=& -\ri \de_{\a}^i \mathbf{W}\non\\
&=& ~~ G^{-1} \Bigg\{ -\frac{\ri}{2} \de_{\a \b} \vf^{\b i}
+ \frac{3\ri}{4} W_{\a \b} \vf^{\b i} + \frac{3\ri}{8} G^{ij} X_{\a j}\Bigg\}
\non\\
&&
+ G^{-3} \Bigg\{ -\frac{\ri}{8} F G^{ij} \vf_{\a j} - \frac{\ri}{8} G^{ij} \cH_{\a \b} \vf^{\b}_{j}
+ \frac{\ri}{4} G_{jk} \vf^{\b k} \de_{\a \b} G^{ij}
+ \frac{1}{4} \vf^{\b i} \vf_{\b}^j \vf_{\a j} \Bigg\}
\non\\
&&
+  G^{-5} \Bigg\{ -\frac{3}{8} G^{ij} G_{kl} \vf^{\b k} \vf_{\b}^l \vf_{\a j}\Bigg\}~,
\\
%%%%%%%%%%%%%%%%%%%%%%%%%%%%%%%%%%%%%%%%%%%%%%%%%%
\mathbf{X}^{ij} 
&=& \frac{\ri}{4} \nabla^{\hal (i} \nabla_\hal^{j)} \mathbf{W} 
\non\\
&=& ~~ G^{-1} \Bigg\{\, \hf \Box G^{ij} 
+ \frac{3}{64} W^{ab} W_{ab} G^{ij} -\frac{3}{64} Y G^{ij}
+ \frac{3\ri}{4} X^{\a (i} \vf_{\a}^{j)}
\Bigg\}\non\\
&&+ G^{-3} \Bigg\{ -\frac{1}{16} F^2 G^{ij} - \frac{1}{16} \cH^{a} \cH_{a} G^{ij} + \frac{1}{4} \cH^{a} G^{k(i} \de_{a} G_k{}^{j)}-\frac{1}{4} G_{kl} \big(\de^{a}   G^{k(i}\big) \de_{a} G^{j)l}
\non\\
&&~~~~~~~~~~
-\frac{3\ri}{8} G^{ij} G_{kl} X^{\a k} \vf_{\a}^l
-\frac{\ri}{8} F \vf^{\a(i} \vf_{\a}^{j)}
+\frac{3\ri}{8} W^{\a \b} G^{ij} \vf^{k}_{\a} \vf_{\b k}
\non\\
&&~~~~~~~~~~
+ \frac{\ri}{16}(\G^a)^{\a \b} \Big(\cH_{a} \vf^{(i}_{\a} \vf^{j)}_{\b} 
+ 8 G^{k(i}  \big(\de_{a} \vf^{j)}_{\a} \big) \vf_{\b k} + 2 \vf_{\a}^{(i}   \big(\de_{a} G^{j)k} \big) \vf_{\b k} \Big)
\Bigg\}\non\\
&&+ G^{-5} \Bigg\{\,\frac{3\ri}{16} F G^{ij} G_{kl} \vf^{\a k} \vf_{\a}^l+ \frac{3\ri}{16} G^{k(i} G^{j)l} (\G^a)_{\a \b} \cH_{a} \vf_{k}^{\a} \vf_{l}^{\b}
-\frac{3\ri}{8}G^{mn}G^{k(i}  \big(\de_{\a \b} G_m{}^{j)} \big) \vf^{\a}_{k} \vf^{\b}_{n}
\non\\
&&~~~~~~~~~
+ \frac{3}{8} G^{k(i} \vf^{j)}_{\a}  \vf^{\a l} \vf^{\b}_{k} \vf_{\b l} -\frac{3}{8} G^{kl} \vf^{\a(i} \vf_{\a}^{j)} \vf_{k}^{\b} \vf_{\b l}
\Bigg\}
\non\\
&&
+  G^{-7} \Bigg\{ \frac{15}{32} G^{ij}G_{kl} G_{mn} \vf^{\a k} \vf_{\a}^{l} \vf^{\b m} \vf_{\b}^n \Bigg\} ~,
\\
%%%%%%%%%%%%%%%%%%%%%%%%%%%%%%%%%%%%%%%%%%%%%%%%%%
\mathbf{F}_{{a} {b}} &=& \frac{1}{4} (\Sigma_{{a} {b}})^{\a \b} \de^{k}_{(\a} \bm{\l}_{\b) k} - W_{{a} {b}} \mathbf{W} \non\\
&=& ~~ G^{-1} \Bigg\{ \hf \de_{[a} \cH_{b]} -\frac{3\ri}{8} G_{ij}X_{ab}{}^{ij} + \frac{\ri}{4} W_{ab \a}{}^i\vf^{\a}_i
\Bigg\} \non\\
&&+ G^{-3} \Bigg\{ \,\frac{1}{4} G_{ij} \cH_{[a} \de_{b]} G^{ij}
-\frac{1}{4} G_{ij}  \big(\de_{[a} G^{ik} \big) \de_{b]} G_k{}^{j} + \frac{\ri}{2} G_{ij} (\G_{[a})^{\a \b} (\de_{b]} \vf_{\a}^i) \vf_{\b}^{j} 
\non\\
&&
~~~~~~~~~
- \frac{\ri}{8} (\G_{[a})^{\a \b}  \big(\de_{b]} G_{ij} \big) \vf^{i}_{\a} \vf^{j}_{\b}
\Bigg\}
\non\\
&&
+ G^{-5}  \Bigg\{ -\frac{3\ri}{8} G^{k}{}_{(i} G^{l}{}_{j)} (\G_{[a})^{\a \b}  \big(\de_{b]} G_{kl} \big) \vf_{\a}^{i} \vf^{j}_{\b} \Bigg\}
~.
\eea
\esubeq

\subsection{Gauged supergravity action}

An off-shell formulation for 5D minimal supergravity can be obtained by coupling the standard Weyl multiplet to two off-shell compensators: vector and linear multiplets \cite{Zucker1,Zucker2,Zucker3,Ohashi1,Ohashi2,Ohashi3,Ohashi4,Bergshoeff1,Bergshoeff2,Bergshoeff3,BKNT-M14}.
This is the 5D analogue of the off-shell formulation for 4D $\cN=2$ supergravity \cite{deWit:1982na, K-08}.
The complete (gauged) supergravity action, $S_{\rm gSG}$, is given by the following two-derivative action:
\begin{subequations} \label{gsg-action}
\bea
S_{\rm gSG}&=&  S_{\rm VM} + S_{\rm L} +\k \, S_{\rm BF} =
 \int \rd^{5|8}z\, E\, \Big\{ 
\frac{1}{4} V_{ij} { H }_{\rm VM}^{ij} 
+\O \mathbf{ W } 
+\k V_{ij} G^{ij} \Big\} \\
&=&
 \int \rd^{5|8}z\, E\,  \Big\{ 
\frac{1}{4} V_{ij} { H }_{\rm VM}^{ij} 
+ \O \mathbf{ W}  
+\k \O W \Big\}~. \label{rep8.13b}
\eea
\end{subequations}
The BF action $\k S_{\rm BF}$ describes a supersymmetric cosmological term. The case $\k=0$ case corresponds to Poincar\'e supergravity, while $\k \neq 0$ leads to gauged or anti-de Sitter supergravity. 

Upon gauge fixing dilatation and superconformal symmetries (dilatation, $S$ and $K$) and integrating out the various auxiliary fields, one obtains the on-shell Poincare supergravity action of \cite{Cremmer, CN}. The contributions from the scalar curvature terms in eqs.~\eqref{bosonic-swm} and \eqref{eq:TensorComp} combine to give the normalised Einstein-Hilbert term $-\hf \cR$ plus a cosmological constant, see, e.g., \cite{BKNT-M14} for details. 

In the remaining subsections, we elaborate on the structure of three independent curvature-squared invariants \cite{BKNT-M14, HOT, BRS, OP1, OP2}. These invariants were constructed in superspace \cite{BKNT-M14} in the standard Weyl multiplet background. In particular, we present the full expressions of all the composite primary multiplets which generate these invariants with the $\log$ multiplet appearing  for the first time in its expanded form in terms of the descendants of $W$ and $W_{\a\b}$.

\subsection{Weyl-squared}
\label{weyl_squared}

We first consider a composite primary superfield that may be used to generate a supersymmetric completion of a Weyl-squared term. In superspace, it was described in \cite{BKNT-M14} in terms of the super Weyl
tensor:
\begin{equation}
    H^{i j}_{\textrm{Weyl}} := - \frac{\ri}{2}   W^{{\alpha} {\beta} {\gamma} i} W_{{\alpha} {\beta} {\gamma}}\,^{j}+\frac{3\ri}{2}   W^{{\alpha} {\beta}} X_{{\alpha} {\beta}}\,^{i j} - \frac{3 \ri}{4}   X^{{\alpha} i} X^{j}_{{\alpha}}~.
    \label{H-Weyl}
\end{equation}
It can be checked that $H^{ij}_{\rm Weyl}$ satisfies the constraints \eqref{linear-constr}.
The superfield $H^{ij}_{\rm Weyl}$ corresponds to the composite linear multiplet first constructed in components by Hanaki, Ohashi, and Tachikawa in \cite{HOT}.

With the aid of the relations \eqref{eq:Wdervs}, the component fields of the composite linear multiplet are straightforward to compute. They include the $\q =0$ projection (or the ``bar-projection'') of $H^{ij}_{\rm Weyl}$,  together with the bar-projection of the following descendant superfields of $H^{ij}_{\rm Weyl}$:
\bsubeq \label{desc-H-Weyl}
\begin{align}
\varphi_{\rm Weyl}^{\hal\, i} &= \frac{1}{3} \nabla^{\hal}_{j} H^{ij}_{\rm Weyl} \ , \\
 F_{{\rm Weyl}} &=  \frac{\ri}{12} \de^{\a}_{i} \de_{\a j}H^{i j}_{\rm{Weyl}}~, \\
\cH^{{a}}_{{\rm Weyl}} &= \frac{\ri}{12} (\Gamma^{{a}})^{\a \b} \de_{\a i} \de_{\b j} H^{i j}_{\rm{Weyl}}~.
\end{align}
\esubeq
Eqs.~\eqref{desc-H-Weyl} play an important role in analysing superconformal primary equations of motion in the next section.  The resulting expression coincides, up to notations, to the results of \cite{HOT}. We will give the full component expressions \eqref{desc-H-Weyl} in a follow-up paper. 

By inserting the components of the composite linear multiplet \eqref{H-Weyl} and \eqref{desc-H-Weyl} into the BF action  \eqref{BF-action-0}, one may construct the following higher-derivative invariant in a standard Weyl multiplet background  \cite{BKNT-M14}
\begin{subequations} 
\label{S-Weyl-0}
\bea 
S_{\rm{Weyl}} &=& \int\rd^{5|8}z\, E\,   V_{ij} { H }_{\rm Weyl}^{ij}  \label{S-Weyl} \\
&=&  - \int \rd^5 x \, e \, \bigg( v_{{a}} \cH^{{a}}_{\rm Weyl} + W F_{\rm Weyl} + X_{i j} H^{i j}_{\rm Weyl} + 2 \l^k \varphi_{k \, \rm Weyl} \non\\
   & &~~~
   - \psi_{{a} i} \Gamma^{{a}} \varphi_{\rm Weyl}^i W - \ri \psi_{{a} i} \Gamma^{{a}} \l_{j} H^{i j}_{\rm Weyl} +  \ri \psi_{{a} i} \Sigma^{{a} {b}} \psi_{{b} j} W H^{i j}_{\rm Weyl} \bigg)~,
   \label{S-Weyl-b}
\eea
\end{subequations}
where the spinor indices here are suppressed. This defines  a locally supersymmetric extension of the Weyl squared term \cite{HOT, BRS, OP1, OP2}.

\subsection{$\log{W}$}
\label{logW}

We now consider a composite linear superfield which includes a supersymmetric Ricci tensor-squared term. In superspace, it was described for the first time in \cite{BKNT-M14} in analogy with the construction of a higher-derivative chiral invariant in 4D $\cN=2$ supergravity \cite{Butter:2013lta}. The composite superfield makes use of the standard Weyl multiplet coupled to the off-shell
vector multiplet compensator. It takes the form\footnote{Note that there is an overall minus sign difference between the definition of the $\log{W}$ invariant in this paper and the one of \cite{BKNT-M14}.}
\begin{equation}
    H^{i j}_{{\log{W}}} = - \frac{3 \ri}{40} \Delta^{i j k l}\de_{k l} \log{W} =  \frac{3 \ri}{1280} \de^{(i j} \de^{k l)} \de_{k l} \log{W}~.
    \label{logW-0}
\end{equation}
In general such a linear multiplet could be defined by replacing $W$ with any primary scalar superfield of weight $q$ for which it is possible to prove that \eqref{logW-0} satisfies all the linear multiplet constraints, eq.~\eqref{linear-constr}, see \cite{BKNT-M14}. However, for various applications, we choose to construct it in terms of the vector multiplet superfield strength, $W$.
Due to the complexity in computing the action of six spinor derivatives on the ``log multiplet,'' the component analysis of $H^{ij}_{{\log{W}}}$ has not
appeared so far. 
This calculation can be performed with the aid of the \textit{Cadabra} software.
Here we find that the full expression of $H^{ij}_{{\log{W}}}$ in terms of the descendant superfields of the vector and standard Weyl multiplets is given by
\bea  \label{HlogW-full}
    H^{i j}_{{\log{W}}}
    &=&
    {}  - \frac{3\ri}{8}  W_{{a} {b}}  X^{{a} {b} i j} + \frac{51\ri}{64}  X^{{\alpha} i} X^{j}_{{\alpha}} 
    \non\\
    &&
    +{{W}}^{-1} \Bigg\{\,\frac{9}{64}X^{i j} Y - \frac{3 \ri}{8}  F_{{a} {b}} X^{{a} {b} i j}  - \frac{1}{2}  \Box{{X^{i j}}} - \frac{9}{64}X^{i j} W^{{a} {b}} W_{{a} {b}} + \frac{1}{4} W^{{a}{b}} W_{{a}{b}}{}^{\alpha (i} \lambda^{j)}_{\alpha} \non\\
    &&~~~~~~~~~~
    - \frac{3}{4}(\Gamma^{{a}})^{{\alpha} {\beta}}   X^{(i}_{{\alpha}}  {\nabla}_{{a}}{\lambda^{j)}_{{\beta}}} +\frac{3}{4}(\Gamma^{{a}})^{{\alpha} {\beta}}   \lambda^{(i}_{{\alpha}} {\nabla}_{{a}}{X^{j)}_{{\beta}}}\Bigg\}
    \non\\
    &&
    + {{W}}^{-2} \Bigg\{\,\frac{1}{2} X^{i j}  \Box{{{W}}}+\frac{1}{2}  \big({\nabla}^{{a}}{{W}} \big) {\nabla}_{{a}}{X^{i j}} 
+ \frac{1}{4} F^{{a}{b}} W_{{a}{b}}{}^{\alpha (i} \lambda^{j)}_{\alpha}-\frac{\ri}{2}  \lambda^{{\alpha}(j}  \Box{{\lambda^{i)}_{{\alpha}}}} \non\\
&&~~~~~~~~~~~
-\frac{\ri}{4}    \big({\nabla}^{{a}}{\lambda^{{\alpha} i}} \big) {\nabla}_{{a}}{\lambda^{j}_{{\alpha}}} - \frac{3 \ri}{16} \epsilon^{{a} {b} {c} {d} {e}} (\Sigma_{{a} {b}})^{{\alpha} {\beta}}   W_{{d} {e}} \lambda^{(i}_{{\alpha}}  {\nabla}_{{c}}{\lambda^{j)}_{{\beta}}} - \frac{3 \ri}{8} (\Gamma^{{a}})^{{\alpha} {\beta}}  \lambda^{i}_{{\alpha}} \lambda^{j}_{{\beta}}  {\nabla}^{{c}}{W_{{a} {c}}} \non\\
&&~~~~~~~~~~~
 + \frac{3 \ri}{64}  Y \lambda^{{\alpha} i} \lambda^{j}_{{\alpha}}
+\frac{3}{8}(\Sigma_{{a} {b}})^{{\alpha} {\beta}}  F^{{a} {b}} X^{(i}_{{\alpha}} \lambda^{j)}_{{\beta}}  
+ \frac{3 \ri}{128} W^{{a} {b}} W_{{a} {b}} \lambda^{{\alpha} i} \lambda^{j}_{{\alpha}} 
\non\\
&&~~~~~~~~~~~
+\frac{9 \ri}{256} \epsilon^{{a} {b} {c} {d} {e}} (\Gamma_{a})^{{\alpha} {\beta}}   W_{{b} {c}} W_{{d} {e}} \lambda^{i}_{{\alpha}} \lambda^{j}_{{\beta}} 
- \frac{3}{8} X^{i j} X^{{\alpha} k} \lambda_{{\alpha} k}
\Bigg\} 
\non\\
&&
+ {{W}}^{-3} \Bigg\{\,\frac{1}{8}X^{i j} F^{{a} {b}} F_{{a} {b}}  - \frac{1}{8} X^{i j} X^{k l} X_{k l} - \frac{1}{4} X^{i j}   \big({\nabla}^{{a}}{{W}} \big) {\nabla}_{{a}}{{W}}  \non\\
&&~~~~~~~~~~
- \frac{\ri}{8} \epsilon^{{a} {b} {c} {d} {e}} (\Sigma_{{a} {b}})^{{\alpha} {\beta}}   F_{{d} {e}} \lambda^{(i}_{{\alpha}}  {\nabla}_{c}{\lambda^{j)}_{{\beta}}} - \frac{\ri}{4} (\Gamma^{{a}})^{{\alpha} {\beta}}  F_{{a} {b}} \lambda^{(i}_{{\alpha}}  {\nabla}^{{b}}{\lambda^{j)}_{{\beta}}}  
\non\\
&&~~~~~~~~~~
- \frac{\ri}{4} (\Gamma^{{a}})^{{\alpha} {\beta}}   \lambda^{i}_{{\alpha}} \lambda^{j}_{{\beta}}  {\nabla}^{{c}}{F_{{a} {c}}} 
+\frac{\ri}{4} (\Gamma^{{a}})^{{\alpha} {\beta}}  X^{i j} \lambda^{k}_{{\alpha}} {\nabla}_{{a}}{\lambda_{{\beta} k}}+\frac{\ri}{4} (\Gamma^{{a}})^{{\alpha} {\beta}}  X^{k (i} \lambda^{j)}_{{\alpha}}  {\nabla}_{{a}}{\lambda_{{\beta} k}} 
\non\\
&&~~~~~~~~~~
 - \frac{\ri}{4} (\Gamma^{{a}})^{{\alpha} {\beta}}  \lambda^{(i}_{{\alpha}}    \big({\nabla}_{{a}}{X^{j) l}} \big) \lambda_{{\beta} l}
 + \frac{\ri}{4}   \lambda^{{\alpha} i} \lambda^{j}_{{\alpha}}  \Box{{{W}}} + \frac{3 \ri}{4}   \big({\nabla}^{{a}}{{W}} \big) \lambda^{{\alpha} (i}   {\nabla}_{{a}}{\lambda^{j)}_{{\a}}} 
 \non\\
&&~~~~~~~~~~
- \frac{\ri}{2} (\Sigma_{{a} {b}})^{{\alpha} {\beta}}  \big({\nabla}^{{a}}{{W}}  \big)\lambda^{(i}_{{\alpha}}   {\nabla}^{{b}}{\lambda^{j)}_{{\beta}}}  
 - \frac{3 \ri}{16} (\Gamma^{{a}})^{{\alpha} {\beta}}  W_{{a} {b}} \lambda^{i}_{{\alpha}} \lambda^{j}_{{\beta}}  {\nabla}^{{b}}{{W}} + \frac{3 \ri}{32} W_{{a} {b}} F^{{a} {b}} \lambda^{{\alpha} i} \lambda^{j}_{{\alpha}} 
 \non\\
 &&~~~~~~~~~~
+\frac{9 \ri}{64} \epsilon^{{a} {b} {c} {d} {e}} (\Gamma_{{a}})^{{\alpha} {\beta}}   W_{{b} {c}} F_{{d} {e}} \lambda^{i}_{{\alpha}} \lambda^{j}_{{\beta}}  - \frac{3 \ri}{32}  (\Sigma_{{a} {b}})^{{\alpha} {\beta}}  X^{i j} W^{{a} {b}} \lambda^{k}_{{\alpha}} \lambda_{{\beta}k}  
 \non\\
 &&~~~~~~~~~~
- \frac{3 \ri}{8}  X^{{\alpha} k} \lambda^{(i}_{{\alpha}} \lambda^{{\beta} j)} \lambda_{{\beta} k}- \frac{3 \ri}{8}   X^{{\alpha} k} \lambda^{{\beta} i} \lambda^{j}_{{\beta}} \lambda_{{\alpha} k} \Bigg \}
 \non\\
 &&
 + {{W}}^{-4}\Bigg\{  -\frac{3 \ri}{16}   \lambda^{{\alpha} i} \lambda^{j}_{{\alpha}}   \big({\nabla}^{{a}}{{W}} \big) {\nabla}_{{a}}{{W}} 
 - \frac{3 \ri}{8} (\Gamma^{{a}})^{{\alpha} {\beta}}   X^{ k (i} \lambda^{j)}_{{\alpha}} \lambda_{{\beta}k}  {\nabla}_{{a}}{{W}} \non\\
 &&~~~~~~~~~~~
 +\frac{3\ri}{8} (\Gamma^{{a}})^{{\alpha} {\beta}}  F_{{a} {b}} \lambda^{i}_{{\alpha}} \lambda^{j}_{{\beta}}  {\nabla}^{{b}}{{W}} 
 + \frac{3 \ri}{32} F^{{a} {b}} F_{{a} {b}} \lambda^{{\alpha} i} \lambda^{j}_{{\alpha}} +\frac{3\ri }{64} \epsilon^{{a} {b} {c} {d} {e}} (\Gamma_{{a}})^{{\alpha} {\beta}}   F_{{b} {c}} F_{{d} {e}} \lambda^{i}_{{\alpha}} \lambda^{j}_{{\beta}}  \non\\
 &&~~~~~~~~~~~
 - \frac{3 \ri}{16} (\Sigma_{{a} {b}})^{{\alpha} {\beta}}  X^{i j} F^{{a} {b}} \lambda^{k}_{{\alpha}} \lambda_{{\beta} k}    -\frac{3 \ri}{16}  X^{i j} X^{k l} \lambda^{\alpha}_{k} \lambda_{{\alpha} l}   -\frac{3 \ri}{32}  X^{k l} X_{ k l} \lambda^{{\alpha} i} \lambda^{j}_{{\alpha}}  \non\\
 &&~~~~~~~~~~~
 -\frac{15}{64} (\Gamma^{{a}})^{{\alpha} {\beta}}   \lambda^{i}_{{\alpha}} \lambda^{j}_{{\beta}} \lambda^{{\gamma} k}  {\nabla}_{{a}}{\lambda_{{\gamma} k}}-\frac{9}{32} (\Gamma^{{a}})^{{\alpha} {\beta}}    \lambda^{(i}_{{\alpha}} \lambda^{{\rho} j)} \lambda^{k}_{{\beta}}  {\nabla}_{{a}}{\lambda_{{\rho} k}}
 \non\\
 &&~~~~~~~~~~~
 -\frac{15}{32}(\Gamma^{{a}})^{{\alpha} {\beta}}   \lambda^{(i}_{{\alpha}} \lambda^{{\rho} j)} \lambda^{k}_{{\rho}}  {\nabla}_{{a}}{\lambda_{{\beta} k}}
   -\frac{15}{64} (\Gamma^{{a}})^{{\alpha} {\beta}}   \lambda^{{\rho} i} \lambda^{j}_{{\rho}} \lambda^{k}_{{\alpha}}  {\nabla}_{{a}}{\lambda_{{\beta} k}}  
      \non\\
   &&~~~~~~~~~~~
+ \frac{9}{32} (\Sigma_{{a} {b}})^{{\alpha} {\beta}}   W^{{a} {b}} \lambda^{(i}_{{\alpha}} \lambda^{{\rho} j)} \lambda^{k}_{{\beta}} \lambda_{{\rho} k} 
+ \frac{9}{64} (\Sigma_{{a} {b}})^{{\alpha} {\beta}}  W^{{a} {b}} \lambda^{{\rho} i} \lambda^{j}_{{\rho}} \lambda^{k}_{{\alpha}} \lambda_{{\beta} k} \Bigg\} 
   \non\\
   &&
   + {{W}}^{-5} \Bigg\{\, \frac{3}{8} (\Gamma^{{a}})^{{\alpha} {\beta}}    \lambda^{(i}_{{\alpha}} \lambda^{{\rho} j)} \lambda^{k}_{{\beta}} \lambda_{{\rho} k}  {\nabla}_{{a}}{{W}}  + \frac{3}{8} (\Sigma_{{a} {b}})^{{\alpha} {\beta}}   F^{{a} {b}} \lambda^{{\rho} i} \lambda^{j}_{{\rho}} \lambda^{k}_{{\alpha}} \lambda_{{\beta} k}    \non\\
   &&~~~~~~~~~~
  +\frac{3}{8}  X^{k l} \lambda^{{\alpha} i} \lambda^{j}_{{\alpha}} \lambda^{{\beta}}_{k} \lambda_{{\beta} l} 
  \Bigg\} \non\\
  &&
  + {W}^{-6}\Bigg\{\,\frac{3 \ri}{32}   \lambda^{{\alpha} i} \lambda^{j}_{{\alpha}} \lambda^{{\beta} k} \lambda^{l}_{{\beta}} \lambda_{k}^{{\gamma}} \lambda_{{\gamma} l} 
  +\frac{3 \ri}{16}   \lambda^{{\alpha} i} 
  \lambda^{k}_{{\alpha}}  
  \lambda^{{\beta} j}
  \lambda^{l}_{{\beta}} 
  \lambda_{k}^{{\gamma}}
  \lambda_{{\gamma} l} \Bigg\} 
~.
\eea
Using the explicit expression \eqref{HlogW-full}, 
together with the relations \eqref{eq:Wdervs}, \eqref{S-on-X_Y-a}, \eqref{VMIdentities}, and \eqref{S-on-W},
we have shown that $H^{i j}_{{\log{W}}}$ is indeed a primary and linear superfield satisfying \eqref{linear-constr}.
Furthermore, we have computed the descendants of the primary superfield $H^{i j}_{{\log{W}}}$ defined as
\bsubeq \label{desc-H-log}
\begin{align}
\varphi_{{\log{W}}}^{\hal\, i} &= \frac{1}{3} \nabla^{\hal}_{j} H^{ij}_{{\log{W}}} \ , \\
 F_{{{\log{W}}}} &=  \frac{\ri}{12} \de^{\a}_{i} \de_{\a j}H^{i j}_{\rm{log W}}~, \\
\cH^{{a}}_{{{\log{W}}}} &= \frac{\ri}{12} (\Gamma^{{a}})^{\a \b} \de_{\a i} \de_{\b j} H^{i j}_{\rm{log W}}~.
\end{align}
\esubeq
By using \eqref{HlogW-full} and the BF action, eqs.~\eqref{SGV} and \eqref{BF-Scomp}, one may construct the following locally superconformal invariant in a standard Weyl multiplet background
\begin{subequations} 
\label{S-log-0}
\bea
S_{\log{W}} &=& \int\rd^{5|8}z\, E\,   V_{ij} { H }_{{\log{W}}}^{ij} \label{S-log} \\
&=&  - \int \rd^5 x \, e \, \bigg( v_{{a}} \cH^{{a}}_{{\log{W}}} + W F_{{\log{W}}} + X_{i j} H^{i j}_{{\log{W}}} + 2 \l^k \varphi_{k \, {\log{W}}}  \non\\
    &&- \psi_{{a} i} \Gamma^{{a}} \varphi_{{\log{W}}}^i W - \ri \psi_{{a} i} \Gamma^{{a}} \l_{j} H^{i j}_{{\log{W}}} +  \ri \psi_{{a} i} \Sigma^{{a} {b}} \psi_{{b} j} W H^{i j}_{{\log{W}}} \bigg)~. \label{S-log-b}
\eea
\end{subequations}
The resulting component action includes, for example, a $(\Box\Box W)$ term which, upon gauge-fixing $W=1$, includes a Ricci tensor squared combination.
A more detailed discussion of the (fairly involved) component structure of \eqref{desc-H-log} will be given elsewhere.

\subsection{Scalar curvature squared} \label{scalar_squared}

Given a composite vector multiplet \eqref{comp-W} and its corresponding descendants \eqref{desc-W-R2}, 
we can then construct a composite linear multiplet defined by \cite{BKNT-M14}
\bea
H^{ij}_{R^2}:=H^{ij}_{\rm{VM}}[\mathbf{W}] &=& \ri (\de^{\a(i} \mathbf{W}) \de_{\a}^{j)} \mathbf{W} + \frac{\ri}{2} \mathbf{W} \de^{\a(i} \de_{\a}^{j)} \mathbf{W} \non\\
&=& -\ri \bm{\lambda}^{\a i} \bm{\l}_{\a}^j +2 \mathbf{W} \mathbf{X}^{ij}~.
\eea
Inserting the composite field $H^{ij}_{R^2}$ and its independent descendants ($\varphi^{\a i}_{R^2}, F_{R^2}$, and $\cH^{{a}}_{R^2}$) into the BF action principle \eqref{BF-action-0} leads to the 
following supersymmetric invariant 
\bsubeq \label{action-R2}
\bea
S_{R^2} &=& \int\rd^{5|8}z\, E\,   V_{ij} H^{ij}_{R^2} \\
&=&  - \int \rd^5 x \, e \, \bigg( v_{{a}} \cH^{{a}}_{R^2} + W F_{R^2} + X_{i j} H^{i j}_{R^2} + 2 \l^k \varphi_{k \, R^2} \non\\
    &&~~~~~~
    - \psi_{{a} i} \Gamma^{{a}} \varphi_{R^2}^i W - \ri \psi_{{a} i} \Gamma^{{a}} \l_{j} H^{i j}_{R^2} +  \ri \psi_{{a} i} \Sigma^{{a} {b}} \psi_{{b} j} W H^{i j}_{R^2} \bigg)~. \label{action-R2-b}
\eea
\esubeq
At the component level, the above action generates the scalar curvature-squared invariant constructed in \cite{OP1,OP2}.

\section{Superconformal equations of motion}
\label{5DEOMSuperspace}

Let us now combine the gauged supergravity action, $S_{\rm gSG}$, with the three independent curvature-squared invariants described by eqs. \eqref{S-Weyl-0}, \eqref{S-log-0}, and \eqref{action-R2}, to form a higher-derivative action 
\bea
S_{\rm HD}&=&  S_{\rm gSG} + \a \, S_{\rm Weyl} +\b  \, S_{{\log{W}}} + \g \, S_{R^2}
~.
\label{HD-action}
\eea
The goal of this section is to obtain superconformal primary equations of motion in superspace that describe minimal 5D gauged supergravity deformed by an arbitrary combination of the three curvature-squared invariants described by the action above.   

We can obtain these equations of motion by varying the superspace action \eqref{HD-action} with respect to the superfield prepotentials of the standard Weyl multiplet ($\mathfrak U$), the vector multiplet compensator ($V_{ij}$), and the linear multiplet compensator ($\Omega$). Such variations lead to the supercurrent superfield $\cJ$, the linear multiplet of the EOM of $V_{ij}$, and the vector multiplet of the EOM of $\Omega$, respectively.
Alternatively, we can reduce \eqref{HD-action} to components and vary it  with respect to the highest dimension independent fields ($Y$, $X_{i j}$, and $F$) of the corresponding multiplets.  The resulting equations of motion then describe the primary fields, i.e., the bottom components, of the multiplets of the equations of motion that arise from the variation of the full superfields. It is then straightforward to reinterpret them as the primary superfields of the equations of motion. By making use of code developed in \textit{Cadabra}, the full higher-derivative action in components has been obtained by substituting the explicit form of composite primary multiplets described in subsections \ref{weyl_squared} to \ref{scalar_squared} together with their descendants. These results, and details of the derivation of the equations of motion which we derived by using a combination of both superspace and components arguments, will be presented in an upcoming paper. The important point to stress is that the equations of motion are fully locally superconformal covariant. From them, successively acting with spinor derivatives (which is equivalent to taking successive $Q$-supersymmetry transformations) one can obtain the whole tower of independent equations of motion. Note that the component action computed from \eqref{HD-action} includes thousands of terms when fermions are considered, and it is not manifestly covariant due to the presence of naked gravitini and Chern-Simons terms. These would become covariant only after taking field variations and several integration by parts. An efficient alternative, and algorithmic, way to attack this problem is then by analysing the multiplets of the equations of motion starting from their primaries. Moreover, one could extract as much information as possible about the structure of the on-shell action (including all fermionic contributions) by directly working with the equations of motion in superspace. 

In the next subsections, we will simply state the final results and show that the three primary equations of motion satisfy all necessary consistency checks dictated by their general structures. From this point of view, the results of this section stand on their own.

\subsection{Vector multiplet}

The vector multiplet equation of motion is obtained by varying \eqref{HD-action} with respect to the superfield $V_{ij}$ or equivalently the field $X_{ij}$. 
The resulting EOM is
\bea
0&=&
\frac{3}{4}H^{ij}_{\rm VM} 
+ \k G^{ij}
+\alpha H^{ij}_{\rm Weyl} 
+ \beta H^{ij}_{\log{W}}
+\gamma H^{i j}_{R^2}
  ~. \label{VM-eom}
\eea
Note that the first two terms correspond to the EOM for the vector multiplet in the two-derivative supergravity theory $S_{\rm gSG}$ while the remaining three terms describe the contribution coming from the three curvature-squared invariants. It is clear that, as expected, the right-hand side of \eqref{VM-eom} is a linear multiplet satisfying \eqref{linear-constr}.

\subsection{Linear multiplet}

The linear multiplet equation of motion is obtained by varying \eqref{HD-action} with respect to the superfield $\Omega$ or equivalently the auxiliary field $F$. 
The resulting EOM is
\be
 0=\mathbf{W} + \kappa W + \gamma W_{R^2}~,
 \label{LM-eom}
\ee
with
\bea
W_{R^2}
&= &
G^{-1} \bigg[~
\hf X^{ij} \mathbf{X}_{ij}
-\frac{1}{2} F^{\ha \hb} \mathbf{F}_{\ha \hb}
+W \Box \mathbf{W} + \mathbf{W} \Box W+  \big(\de^{\ha} W \big) \de_{\ha} \mathbf{W}
\non\\
&&~~~~~~~
+ \frac{3}{16} Y W \mathbf{W}
-\frac{3}{2} W^{\ha \hb} \big(F_{\ha \hb} \mathbf{W} + \mathbf{F}_{\ha \hb} W \big)
-\frac{39}{16} W^{\ha \hb} W_{\ha \hb} W \mathbf{W} 
\non\\
&&~~~~~~~
+ \frac{\ri}{2} \l^{\a i} \de_{\a}{}^{\b} \bm{\l}_{\b i}
+ \frac{\ri}{2} \bm{\l}^{\a i} \de_{\a}{}^{\b} \l_{\b i}
+ \frac{3}{2} X^{\a i} \big( W \bm{\l}_{\a i} + \bm{W} \l_{\a i}\big) 
-\frac{3\ri}{2} W^{\a \b} \l_{\a}^i \bm{\l}_{\b i} 
\bigg]
\non\\
&&
+G^{-3} \bigg[~
\frac{1}{4} G^{ij} \vf_{\b i} \big( \l_{\a j} \de^{\a \b} \mathbf{W} + \bm{\l}_{\a j} \de^{\a \b} W
\big)+ \frac{1}{2} G^{ij} \vf_{\b i} \big( W \de^{\a \b} \bm{\l}_{\a j} + \mathbf{W} \de^{\a \b} \l_{\a j}
\big)
\non\\
&&~~~~~~~~
-\frac{1}{2}G^{ij} \vf_{\a i} \big( F^{\a \b} \bm{\l}_{\b j} + \mathbf{F}^{\a \b} \l_{\b j} \big)
-\frac{1}{4}G^{ij} F \big( W \mathbf{X}_{ij} + \mathbf{W} X_{i j} \big)
\non\\
&&~~~~~~~~
-\frac{3}{4}G^{ij} W^{\a \b} \vf_{\a i} \big(\mathbf{W} \l_{\b j} + W \bm{\l}_{\b j} \big)
+\frac{\ri}{4}F G^{ij}\l^{\a}_{i} \bm{\l}_{\a j}
+ \frac{3\ri}{4} G_{ij} X^{\a i} \vf_{\a}^j W \mathbf{W} 
\non\\
&&~~~~~~~~
+\frac{1}{4} G_{ij} \vf^{\a i} \big( X^{jk} \bm{\l}_{\a k} + \mathbf{X}^{jk} \l_{\a k} \big)
-\frac{\ri}{4} \vf^{\a i} \vf^{j}_\a \big( X_{ij} \mathbf{W} + \mathbf{X}_{ij} W \big)
\non\\
&&~~~~~~~~
- \frac{1}{4} \vf^{\a i} \vf^{j}_{\a} \l^{\b}_i \bm{\l}_{\b j}^L
\bigg]
\non\\
&&
+G^{-5} \bigg[~
\frac{3\ri}{8} G^{ij} G^{kl} \vf^{\a}_{k} \vf_{\a l}
\Big( 
X_{ij} \mathbf{W} + \mathbf{X}_{ij} W 
-\ri \l^{\b}_i \bm{\l}_{\b j}
\Big) \bigg] 
~.
\label{LM-EOM-111}
\eea
It is possible to check explicitly that $W_{R^2}$ is primary, $S_{\a}^i W_{R^2} = 0$. 
Moreover, we find that $W_{R^2}$ can be expressed as
\bsubeq\label{W-R2-composite}
\begin{equation} \label{WR2_as_W1}
    W_{R^2} = \frac{\ri}{32} G \de_{i j} \mathcal{R}^{i j}_{1}~,
\end{equation}
where
\begin{equation}
    \mathcal{R}^{i j}_{1} =G^{-2} \left(\d^i_k \d^j_l - \frac{1}{2 G^2} G^{i j}G_{k l}\right) H^{k l}_{\rm bilinear}
    ~,
\end{equation}
and
\begin{equation}
    H^{k l}_{\rm bilinear} = 2 W \mathbf{X}^{k l} + 2 \mathbf{W} X^{k l} - 2 \l^{\a k} \bm{\l}^{l}_{\a}~.
\end{equation}
\esubeq 
This is exactly the structure of the composite vector multiplets $\mathbb{W}_{n}$ in (\ref{W_family}) with $n=1$ and with a precise choice of composite linear multiplet $H^{k l}:= H^{kl}_{\rm bilinear}$. See section \ref{NewCurvature2} for more detail on these composite vector multiplets. Besides the remarkably simple form of \eqref{W-R2-composite}, this result guarantees that the right-hand side of eq.~\eqref{LM-eom}, and in particular \eqref{LM-EOM-111}, is a primary superfield satisfying the vector multiplet constraints \eqref{vector-defs}, as expected. This is a very non-trivial consistency check of eq.~\eqref{LM-EOM-111}.

\subsection{Standard Weyl multiplet}
The conformal supergravity equation of motion is obtained by varying \eqref{HD-action} with respect to the standard Weyl multiplet prepotential superfield $\mathfrak U$ or equivalently the field $Y$. The resulting EOM is
\bsubeq \label{Y-eom}
\bea
0&=& \cJ=J_{\rm EH}
+\alpha J_{\rm Weyl}
+\beta J_{\log{W}} + \gamma J_{{R^2}}~,
\label{Y-eom-0}
\eea
with
\bea
J_{\rm EH}
&=&
\frac{3}{32}(G- W^{3}) 
~,
\label{Y-eom-1}
\\
J_{\rm Weyl}
&=&
- \frac{3}{64} W Y + \frac{3}{16} W W^{{a} {b}}W_{{a} {b}}  + \frac{3}{32}  F_{{a} {b}} W^{{a} {b}}   - \frac{3}{16} \l_{i}^{\alpha} X^{i}_{\alpha}
~,
\\
J_{\log{W}}
&=&
-\frac{3}{1024}W Y 
-  \frac{69}{1024} W^{{a} {b}} W_{{a} {b}}  W 
+  \frac{3}{32}  \Box W 
-  \frac{3}{64} F_{{a}{b}} W^{{a} {b}}
- \frac{3}{256}\l^{\alpha}_{j} X^{j}_{\alpha} 
\non\\
&&+ \frac{3}{128} F^{{a} {b}} F_{{a} {b}}  W^{-1}  
-  \frac{9}{128}  X^{i j} X_{i j}  W^{-1} 
+  \frac{3\ri}{32}  (\Gamma^{{a}})^{\alpha \beta} W^{-1}  \l^{i}_{\alpha} {\de}_{{a}}\l_{\beta i} 
\non\\
&&+  \frac{3}{64}  W^{-1} ({\de}^{{a}}W) {\de}_{{a}}W  
- \frac{3 \ri }{128} (\Sigma^{{a} {b}})^{\alpha \beta}  F_{{a}{b}} \l^{i}_{\alpha} \l_{\beta i} W^{-2}   
-\frac{3\ri}{64}    X^{i j}  \l^{\alpha}_{i} \l_{ \alpha j} W^{-2}  
\non\\
&&-  \frac{3 \ri}{32}  (\Gamma_{{b}})^{\alpha \beta}  \  \l_{\alpha}^{j} \l_{{\beta} j} W^{-2} {\de}^{{b}} W  
- \frac{3}{256}     \l^{\alpha i} \l^{\beta}_{i} \l^{j}_{\alpha} \l_{{\beta} j} W^{-3}
~,
\label{Y-eom-2}
\\
J_{{R^2}}
&=&
-\frac{3}{8} W \mathbf{W}^2  
+\frac{3}{32} G^{-1} \Big(\,
W G^{ij} \mathbf{X}_{ij} 
+ G^{ij}X_{ij} \mathbf{W}
-\ri G^{ij} \lambda^{\a}_{i} \bm{\l}_{\a j}
\Big)
~.
\label{Y-eom-3}
\eea
\esubeq
Here $J_{\rm EH}$ is the EOM from the gauged supergravity action, $S_{\rm gSG}$, which does not have any contribution from the cosmological constant term $\kappa$.

Analogous to the case of 4D $\cN=2$ conformal supergravity \cite{KT,Butter:2010sc}, the 5D Weyl multiplet may be described by a single unconstrained real prepotential $\mathfrak U$ \cite{BKNT-M14}. Given a system of matter superfields $\vf^i$, one can construct a Noether coupling between $\mathfrak U$ and the matter supercurrent $\cJ$ of the form 
\bea
S[\vf^i] &=& \int\rd^{5|8}z\, E\,   {\mathfrak U} \cJ = \int \rd^5 x \, e \, \Big( Y J + \cdots \Big)~,
\eea
where $J = \cJ \loco$. 
The supercurrent $\cJ$ is a dimension-3 primary real scalar superfield. The conformal supergravity EOM \eqref{Y-eom} is obtained by varying the supergravity action with respect to $\mathfrak U$
\bea
\frac{\d S[\vf^i]}{\d \mathfrak U} = \cJ = 0~.
\eea
The supercurrent multiplet in 5D was constructed by Howe and Lindstr\"om \cite{HL}. It satisfies the conservation equation 
\bea
\de^{ij} \cJ = 0~, \label{supercurrent}
\eea
when all matter superfields equations of motion are satisfied.
Thus, as a consistency check, we shall prove that the expression $\cJ$ in \eqref{Y-eom} satisfies the conservation constraint \eqref{supercurrent}. It has been shown in \cite{BKNT-M14} that this constraint holds for $J_{\rm EH}$. 
For each invariant, we have indeed verified that the corresponding $J$ is a primary superfield of dimension 3. It also satisfies
$
\de^{ij}J=0
$
provided the vector and linear multiplets equations of motion of eqs.~\eqref{VM-eom} and \eqref{LM-eom}, respectively, are imposed. Using \textit{Cadabra} an explicit calculation shows that, off-shell, it holds
\bsubeq \label{supercurrent_equation}
\bea
\de^{ij}J_{\rm Weyl}&=&\frac{3 \ri}{4} W H^{ij}_{\rm Weyl}
~,\\
\de^{ij}J_{\log{W}}&=&\frac{3 \ri}{4} W H^{ij}_{\log{W}}
 ~,\\
\de^{ij}J_{R^2}&=& \frac{3 \ri}{4}  W H^{ij}_{R^2}
- \frac{3 \ri}{4} G^{ij}W_{R^2}
~.
\eea\esubeq 
It is then clear that the right-hand sides of \eqref{supercurrent_equation} are proportional to the composite vector and linear multiplets appearing in \eqref{VM-eom} and \eqref{LM-eom}. Consequently, the supercurrent conservation equation \eqref{supercurrent} is satisfied once the equations of motion for the compensators are used. This represents a very non-trivial consistency check of \eqref{Y-eom-1}--\eqref{Y-eom-3}.

\section{An alternative scalar curvature-squared invariant} 
\label{NewCurvature2}

Recall the action defined in terms of a composite vector multiplet superfield, $\mathbf{W}$, written in eqs.~(\ref{S_boldW}) and (\ref{comp-W}), respectively. There also exists an infinite number of alternative vector multiplets composite of a linear multiplet compensating superfield, $G_{i j}$, and a superfield associated to a primary real $\cO^{(2n)}$ multiplet $H^{i_1\cdots i_{2n}}=H^{(i_1\cdots i_{2n})}$, such that $\de_\a^{(j}H^{i_1\cdots i_{2n})}=0$. In 5D this was constructed in \cite{BKNT-M14} by extending the 4D $\cN=2$ analysis of \cite{BK11}. We refer to \cite{BKNT-M14} for details, including the precise definition and literature on $\cO^{(2n)}$ multiplets, and simply state the final result here.
The following superfields 
\be \label{W_family}
    \mathbb{W}_{n} 
    = \frac{\ri (2n)!}{2^{2n+3} (n+1)! (n-1)!} G \nabla_{i j} \mathcal{R}_n^{i j}~,
\ee
where
\be
    \mathcal{R}^{i j}_{n} =
    G^{-2n}\left(\d^i_k \d^j_l - \frac{1}{2 G^2} G^{i j}G_{k l}\right) H^{k l i_1  \cdots i_{2n-2}} G_{(i_1 i_2} \cdots G_{i_{2n-3} i_{2n-2})}~,
\ee
all satisfy the constraints \eqref{vector-defs} for any positive integer $n$.
In fact, $\mathbb{W}_1$ is precisely the structure seen in $W_{R^2}$ of eq.~(\ref{WR2_as_W1}). By considering $n=2$, and choosing $H^{ijkl}$ to be the square of a  linear multiplet $H^{ij}$ (distinguished from $G^{ij}$), $H^{ijkl}=H^{(ij}H^{kl)}$, we can engineer an alternative scalar curvature-squared invariant. The result is in spirit similar to the scalar curvature-squared invariant engineered for 4D $\cN=2$ in \cite{deWS} and directly related to 5D $\cN=1$ results in \cite{Ozkan:2016csy}.\footnote{GT-M is grateful for discussions with M.\,Ozkan on scalar curvature-squared invariants.}
Let us show how this works.

Consider the $n=2$ composite superfield:
\be
    \mathbb{W}_{2} = \frac{\ri}{32} G \nabla_{i j} \mathcal{R}^{i j}_{2}~,
    \label{W2-0}
\ee
where
\be
    \mathcal{R}^{i j}_{2} = G^{-4}\left(\d^i_k \d^j_l - \frac{1}{2 G^2} G^{i j}G_{k l}\right) H^{(k l} H^{m n)}G_{m n}~.
\ee
By explicitly computing \eqref{W2-0},
we may define $\mathbb{W}_2$ as a linear combination of real functions, $\mathcal{P}_A$ and $\mathcal{P}_{A B}{}^{i j}$, which are themselves comprised of descendants of the linear multiplets:
\be
    \mathbb{W}_{2} = 2 \mathcal{P}_{A} F^{A} + 2 \ri \mathcal{P}_{A B}{}^{i j}\varphi^{\a A}_i \varphi^{B}_{\a j}
    ~.
    \label{W2-2}
\ee
Here the index $A = 1, 2$ indicates the two linear superfields, $G^1_{i j} = G_{i j}$ and $G^{2}_{i j} = H_{i j}$. Note that this is analogous to eq.~(2.5) in \cite{Ozkan:2016csy} with the first $A$ index fixed so that $\mathcal{F}_{A B} \rightarrow \mathcal{P}_{B}$ and with an appropriate normalisation factor added to the second term. All functions, $\mathcal{P}_A$ and $\mathcal{P}_{A B}{}^{i j}$, are defined as follows:
\bsubeq
\bea
    \mathcal{P}_{1} &=& \frac{1}{8} H^2 G^{-3} - \frac{3}{32} \left(G_{k l} H^{k l}\right)^2 G^{-5}~, \\
    \mathcal{P}_{2} &=& \frac{1}{8} \left(G_{k l} H^{k l}\right) G^{-3}~,\\
    \mathcal{P}_{1 1}{}^{i j} &=& -\frac{3}{16}G^{i j} H^2 G^{-5} - \frac{3}{16} \left(G_{k l} H^{k l}\right) H^{i j} G^{-5} + \frac{15}{64} \left( G_{k l} H^{k l}\right)^2 G^{i j} G^{-7}~, \\
    \mathcal{P}_{1 2}{}^{i j} &=& \mathcal{P}_{2 1}{}^{i j} = \frac{1}{8} H^{i j} G^{-3} - \frac{3}{16} \left( G_{k l} H^{k l} \right) G^{i j}  G^{-5} ~,\\
    \mathcal{P}_{2 2}{}^{i j} &=&  \frac{1}{8} G^{i j} G^{-3}~.
    \label{P-e}
\eea
\esubeq
It is then a straightforward exercise to show that functions with two $A$ indices are derivatives of functions with one, that is:
\be
    \mathcal{P}_{A B}{}^{i j} = \frac{\partial \mathcal{P}_{A}}{\partial G^{B}_{i j}}~.
\ee
They also satisfy the following constraints
\be
    \mathcal{P}_{A B}{}^{i j} = \mathcal{P}_{(A B)}{}^{i j}~, \; \; \; \mathcal{P}_{A B}{}^{i j} G^{B}_{j k} = - \frac{1}{2} \d^i_k \mathcal{P}_{A}~.
\ee
Lastly we may define functions of two derivatives on $\mathcal{P}_{A}$,
\be
    \mathcal{P}_{A B C}{}^{i j k l} := \frac{\partial \mathcal{P}_{A B}{}^{i j}}{\partial G^C_{k l}} = \frac{\partial^2 \mathcal{P}_{A}}{\partial G^B_{i j} \partial G^C_{k l}}~,
\ee
which satisfy
\be
    \mathcal{P}_{A B C}{}^{i j k l} = \mathcal{P}_{(A B C)}{}^{i j k l}~, \; \; \; \mathcal{P}_{A B C}{}^{i j k l} \e_{j k} = 0~.
\ee
These are the constraints needed to ensure that $\mathbb{W}_{2}$ in \eqref{W2-2} satisfies \eqref{vector-defs}, which in our case are satisfied by construction.

To engineer the alternative scalar curvature-squared invariant, consider $G_{i j}$ to be a compensator and $H_{i j}$ to be composite of a vector multiplet, which is built out of a single Abelian
vector multiplet, as in eq.~(\ref{HijVM}),
\be
    H^{i j} := H^{ij}_{\rm{VM}}
    ~.
    \label{Hvm-2}
\ee
In finding the $\mathbb{X}^{i j}_2$ descendant of $\mathbb{W}_2$, we are interested in squared contributions of $F_{\rm VM}$ being a descendant field of $H^{i j}_{\rm VM}$. This is apparent from the fact that $F_{\rm VM}$ satisfies
\be
    F_{\rm VM} = 4 W \Box W
    +\cdots
    ~,
\ee
where $\Box W$ gives rise to a $\cR$
contribution. Roughly speaking, by considering \eqref{W2-2}--\eqref{P-e} with the choice \eqref{Hvm-2}, we are squaring the kinetic term of the vector multiplet compensator which, in turn, leads to a scalar curvature squared invariant. In fact, 
if we look at the dimension-2 scalar descendant of $\mathbb{W}_2$ we obtain
\bsubeq
\bea
    \mathbb{X}^{i j}_2 &:=& \frac{\ri}{4} \nabla^{\a (i}\nabla^{j)}_{\a} \mathbb{W}_2 = \frac{1}{8} G^{i j} G^{-3} F^2_{\rm VM} + \cdots = \mathcal{P}_{2 2}{}^{i j} F^2_{\rm VM} + \cdots
    ~,
    \\
        G_{ij}   \mathbb{X}^{i j}_2&=& 4 G^{-1} W^2(\Box W)^2+\cdots~.
        \label{BFR2Vec}
\eea
\esubeq
Specifically, eq.~\eqref{BFR2Vec} is one term in the component action given by the BF action principle. If one proceeds in setting to constants $G$ and $W$, by gauge fixing dilatation and using (two-derivative) equations of motion, we are left with a $\cR^2$ contribution to the four-derivative component action. Although we have not yet analysed in detail the equations of motion and the component structure of this invariant, we expect it might play a role in studying higher-derivative invariants in alternative off-shell superspace settings, as, for example, the recent off-shell supergravity constructed in \cite{Hutomo:2022hdi} by using the variant hyper-dilaton Weyl multiplet of conformal supergravity. We leave for the future more investigations along this line.

%%%%%%%%%%%%%%%%%%%%%%%%%%%%%%%%%%
%%%%%%%%%%%%%%%%%%%%%%%%%%%%%%%%%%
\vspace{0.3cm}
\noindent
{\bf Acknowledgements:}\\
We are grateful to M.\,Ozkan and Y.\,Pang for discussions and collaboration related to this work.
This work is supported by the Australian Research Council (ARC)
Future Fellowship FT180100353 and by the Capacity Building Package of the University
of Queensland.
G.G. and S.K. are supported 
by the postgraduate scholarships 
at the University of Queensland.

%%%%%%%%%%%%%%%%%%%%%%%%%%%%%%%%%%%%%%%%%%%%%%%%%%%%%%
%%%%%%%%%%%%%%%%%%%%%%%%%%%%%%%%%%%%%%%%%%%%%%%%%%%%%%

\appendix

%%%%%%%%%%%%%%%%%%%%%%%%%%%%%%%%%%%%%%%%%%%%%%%%%%%%%%
%%%%%%%%%%%%%%%%%%%%%%%%%%%%%%%%%%%%%%%%%%%%%%%%%%%%%%

\begin{footnotesize}

\end{footnotesize}

\end{document}